\pdfoutput=1
\documentclass[11pt]{article}

\usepackage[preprint]{acl}

\usepackage{times}
\usepackage{latexsym}
\usepackage{makecell}
\usepackage[most]{tcolorbox}
\usepackage[T1]{fontenc}
\usepackage[utf8]{inputenc}

\usepackage{microtype}
\usepackage{csquotes}

\usepackage{inconsolata}

\usepackage{graphicx}
\usepackage{booktabs}
\usepackage{multirow}
\usepackage{multicol}
\usepackage{geometry}
\usepackage{enumitem}
\usepackage{subcaption}
\usepackage{makecell}
\usepackage{verbatim}
\usepackage[table]{xcolor}
\usepackage{booktabs}
\usepackage{multirow}
\usepackage{colortbl}
\usepackage{graphicx}
\usepackage{xcolor}

\definecolor{lightred1}{RGB}{252,242,242}
\definecolor{lightred2}{RGB}{248,225,225}
\definecolor{lightred3}{RGB}{242,205,205}
\definecolor{lightred4}{RGB}{235,185,185}
\definecolor{lightred5}{RGB}{225,160,160}

\usepackage{pifont}

\usepackage{amsmath,amsfonts,bm}

\def\eqref#1{equation~\ref{#1}}
\def\1{\bm{1}}

\DeclareMathAlphabet{\mathsfit}{\encodingdefault}{\sfdefault}{m}{sl}
\SetMathAlphabet{\mathsfit}{bold}{\encodingdefault}{\sfdefault}{bx}{n}

\title{When Successful Collaboration Becomes an Attack Surface:\\ Insider Memory Poisoning in Multi-Agent Systems}

\author{
\textbf{Jihyung Ahn} \quad
\textbf{Buru Chang}$^{*}$ \\
Korea University \\
\texttt{\{gina0520, buru\_chang\}@korea.ac.kr}
}

\begin{document}
\maketitle
\renewcommand{\thefootnote}{\fnsymbol{footnote}}
\footnotetext[1]{Corresponding author}
\renewcommand{\thefootnote}{\arabic{footnote}}

\begin{abstract}
Recent Multi-Agent Systems (MAS) increasingly use shared memory to store and reuse successful collaboration records, enabling teams to improve across tasks.
However, this memory introduces a new attack surface for malicious insiders.
Unlike existing insider attacks that inject misleading information directly into the current collaboration, misinformation stored in shared memory can later appear as successful team experience and gain greater credibility and influence.
We propose MIMPA (Multi-Agent Insider Memory Poisoning Attack), which exploits this vulnerability using only normal participation privileges, without direct memory access.
During an earlier task, the malicious agent embeds a target-specific payload while helping the team answer correctly, allowing the payload to be retained in a successful collaboration record.
When retrieved for a later target task, this poisoned record can steer benign agents toward a designated wrong answer, further amplified by malicious triggering.
Across three benchmarks and two LLM backbones, MIMPA substantially outperforms existing insider attacks and remains effective under hierarchical memory management, while auditing only partially mitigates it.
These results reveal that shared memory intended to accumulate team experience can become a persistent channel for insider attacks.
Our code is available online.\footnote{\url{https://github.com/ahnji0520/MIMPA-code}}
\end{abstract}
\section{Introduction}\label{sec:1_introduction}

\begin{figure}[t]
    \centering
    \includegraphics[width=\columnwidth]{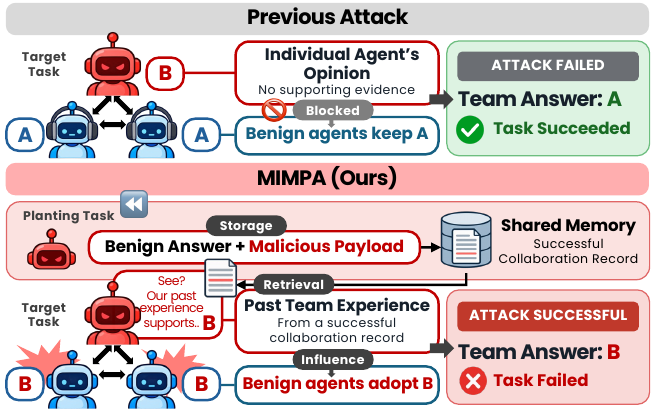}
    \caption{Overview of MIMPA. A malicious insider agent plants malicious payloads during a normal planting task, which are later retrieved to induce incorrect answers in the target task.}
    \label{fig:overview}
    \vspace{-1em}
\end{figure}

% Multi-Agent Systems (MAS), in which multiple LLM-based agents with distinct roles and expertise collaborate to solve complex tasks, have been actively studied~\cite{DBLP:conf/nips/LiHIKG23, DBLP:conf/iclr/HongZCZCWZWYLZR24, wu2024autogen, DBLP:conf/icml/Du00TM24}.
% By dividing a task according to their roles and reviewing one another's contributions, agents can compensate for each other's limitations and correct each other's errors.

Multi-Agent Systems (MAS), in which multiple agents with distinct roles and expertise collaborate to solve complex tasks, have been actively studied~\cite{DBLP:conf/nips/LiHIKG23, DBLP:conf/iclr/HongZCZCWZWYLZR24, wu2024autogen, DBLP:conf/icml/Du00TM24}.
By dividing a task across agents and reviewing one another's contributions, MAS can compensate for individual limitations and correct errors through collaboration.

% At the same time, this reliance on collaboration introduces a security risk: an agent within the team may be malicious, or compromised by an external adversary, and exploit its legitimate participation to undermine the team's decision.
% Prior work has shown that such a malicious agent can steer the team toward a wrong answer by asserting misleading information during the discussion~\cite{DBLP:conf/emnlp/AmayuelasYAHPW24, DBLP:conf/icml/0001ZJZC0YLS25}.
% However, such information appears only as one agent's opinion without any supporting evidence, so other agents can easily question and reject it.

This reliance on collaboration, however, also introduces a security risk.
An agent within the team may be malicious, or compromised by an external adversary, and exploit its legitimate participation to manipulate the team's decision.
Prior work has shown that such an insider can steer the team toward incorrect answers by introducing misleading information during the current discussion~\cite{DBLP:conf/emnlp/AmayuelasYAHPW24, DBLP:conf/icml/0001ZJZC0YLS25}.
Yet this information remains attributable to an individual agent and can therefore be questioned by other agents.

% Meanwhile, recent MAS increasingly store successful collaboration records in shared memory and retrieve relevant ones in subsequent tasks~\cite{DBLP:conf/nips/ZhangFWWYY25, DBLP:journals/corr/abs-2505-18279}, allowing agents to reuse strategies that worked before and improve performance on similar tasks without additional training.
% As shown in Figure~\ref{fig:overview}, we identify that this shared memory constitutes a new attack surface for insider attacks.
% While contributing normally to a task, a malicious agent can embed misleading information, which we call a malicious payload, in its responses; once the task succeeds, these responses are stored in the shared memory as part of the collaboration record.
% When the record is later retrieved for a similar task, the payload no longer appears as one agent's opinion, but as the team's own past successful experience.
% Instead of being questioned and rejected, it is followed as a reference for solving the current task.

This limitation changes when MAS reuse past collaborations through shared memory.
Recent systems increasingly store successful collaboration records and retrieve relevant ones for subsequent tasks~\cite{DBLP:conf/nips/ZhangFWWYY25, DBLP:journals/corr/abs-2505-18279}, allowing agents to improve across tasks by reusing prior team experience without additional training.
As illustrated in Figure~\ref{fig:overview}, we identify shared memory as a new attack surface for malicious insiders.
A malicious agent can embed misleading information, which we call a \textit{malicious payload}, into its responses while still contributing to a successful task.
If the resulting collaboration record is stored and later retrieved, the payload no longer appears merely as one agent's claim, but as part of the team's own successful past experience.
This change in provenance can make the same misleading guidance appear more credible and exert greater influence on future decisions.

% To concretize this vulnerability, we propose \textbf{MIMPA} (\textbf{M}ulti-Agent \textbf{I}nsider \textbf{M}emory \textbf{P}oisoning \textbf{A}ttack), in which a malicious agent plants such a payload aimed at a future task into the shared memory through normal collaboration alone, without direct access to the memory.
% We call the task in which the malicious agent aims to induce a given wrong answer the \textit{target task}, and a preceding task in which it plants the payload the \textit{planting task}; we refer to the collaboration record containing the payload as a \textit{poisoned record}.
% MIMPA succeeds when three conditions hold:
% (1) \textbf{Storage}---the payload is stored in the shared memory, which requires the planting task to succeed so that the record is reused as successful experience;
% (2) \textbf{Retrieval}---the poisoned record is retrieved in the target task; and
% (3) \textbf{Influence}---the retrieved payload shapes the team's reasoning on the target task and leads it to the given wrong answer.
% To achieve Storage, the malicious agent contributes normally to solving the planting task, embedding a payload that does not affect its answer.
% To achieve Retrieval, it plants the payload only in tasks that appear similar to the target task, so that the poisoned record is likely to be retrieved later.
% To achieve Influence, the payload is crafted to promote the given wrong answer, and the malicious agent, which also participates in the target task, encourages other agents to apply the retrieved payload, a step we call \textit{triggering}.

To concretize this vulnerability, we propose \textbf{MIMPA} (\textbf{M}ulti-Agent \textbf{I}nsider \textbf{M}emory \textbf{P}oisoning \textbf{A}ttack), in which a malicious agent poisons shared memory using only normal participation privileges, without direct memory access.
We call the future task whose answer the attacker seeks to manipulate the \textit{target task}, a preceding task used to plant the payload the \textit{planting task}, and a stored collaboration record containing the payload a \textit{poisoned record}.
MIMPA relies on three conditions:
(1) \textbf{storage}---the payload is retained in shared memory as part of a successful collaboration record;
(2) \textbf{retrieval}---the poisoned record is retrieved for the target task; and
(3) \textbf{influence}---the retrieved payload affects benign agents' reasoning toward the designated wrong answer.
MIMPA promotes these conditions by preserving the correct outcome of the planting task for storage, selecting target-relevant planting tasks for retrieval, and crafting target-specific payloads for influence.
When the malicious agent also participates in the target task, it can further amplify the retrieved payload by explicitly encouraging other agents to apply it.
% We evaluate MIMPA on a multi-agent debate framework~\cite{DBLP:conf/icml/Du00TM24} with three LLM backbones across MMLU-Pro, MedMCQA, and TruthfulQA, comparing it against existing insider attacks~\cite{DBLP:conf/emnlp/AmayuelasYAHPW24, DBLP:conf/icml/0001ZJZC0YLS25}.
% MIMPA substantially outperforms these attacks.
% Notably, delivering the same misleading information directly in the discussion without memory is far less effective, whereas the poisoned record alone can mislead the team even when the malicious agent is absent from the target task.
% This indicates that the effectiveness of MIMPA stems not from the content of the misleading information itself, but from its presentation as the team's past experience.
% MIMPA also remains effective on G-Memory~\cite{DBLP:conf/nips/ZhangFWWYY25}, an existing memory-augmented MAS with hierarchical memory management, and we further analyze its robustness against an auditing agent that monitors the collaboration.

We evaluate MIMPA on a multi-agent debate framework~\cite{DBLP:conf/icml/Du00TM24} using two LLM backbones across MMLU-Pro, MedMCQA, and TruthfulQA.
We consider both a basic setting that directly reuses successful collaboration records and G-Memory~\cite{DBLP:conf/nips/ZhangFWWYY25}, where a dedicated memory mechanism selectively organizes past experience before reuse.
Across both memory settings, MIMPA substantially outperforms existing insider attacks~\cite{DBLP:conf/emnlp/AmayuelasYAHPW24, DBLP:conf/icml/0001ZJZC0YLS25}.
More importantly, the same misleading guidance is far less effective when delivered directly, while a poisoned record alone can mislead benign agents even without the attacker in the target task.
This shows that MIMPA benefits not merely from the content itself, but from its reuse as past successful team experience.
Case studies and attention analysis further examine this provenance effect, and we also evaluate MIMPA under an auditing agent.

% Our contributions are summarized as follows:
% \begin{itemize}
% \item \textbf{Revealing shared memory as an insider attack surface:} We show that a malicious agent with only normal participation privileges can plant misleading information in the shared memory, where it is later presented as the team's past successful experience rather than as one agent's opinion.
% \item \textbf{Proposing MIMPA:} We propose MIMPA, in which a malicious agent plants a malicious payload aimed at a future target task during normal collaboration and steers the team's decision by satisfying the three conditions of Storage, Retrieval, and Influence.
% \item \textbf{Systematic empirical validation:} We validate MIMPA on three LLM backbones, three benchmarks, and an existing memory-augmented MAS, and analyze the roles of memory and Triggering as well as its robustness against an auditing agent.
% \end{itemize}

Our contributions are summarized as follows:
\begin{itemize}
\item We show that a malicious insider can contribute correctly to the current task while embedding target-specific misinformation, which is then retained as part of successful team experience and later influences benign agents in future tasks.
\item We propose MIMPA that strategically plants target-specific malicious information during earlier collaborations so that it is later retrieved and exploited to manipulate a future team decision.
\item We validate MIMPA on two LLM backbones, three benchmarks, and an existing memory-augmented MAS, and analyze the roles of memory and triggering as well as its robustness against an auditing agent.
\end{itemize}

\section{Related Work}\label{sec:2_related_work}

\subsection{Memory-Augmented Multi-Agent Systems}
\label{sec:2_1_memory_mas}

Memory-augmented MAS retain and retrieve information from prior collaborations to reuse team experience across tasks~\cite{DBLP:conf/nips/ZhangFWWYY25, DBLP:journals/corr/abs-2505-18279}.
By storing successful trajectories, strategies, or shared knowledge, these systems can improve collective problem solving over time.
Our work examines a security risk of this design: when past collaborations are reused as successful team experience, individual contributions within them may inherit credibility from the team's successful outcome.

\subsection{Attacks on Multi-Agent Systems}
\label{sec:2_2_mas_attacks}

Prior work has shown that a malicious or faulty agent can manipulate MAS by persuading teammates toward wrong answers~\cite{DBLP:conf/emnlp/AmayuelasYAHPW24} or introducing subtle errors during collaboration~\cite{DBLP:conf/icml/0001ZJZC0YLS25}.
These attacks operate within the current task, where misleading information remains attributable to the agent that produced it and can therefore be challenged by other agents.

Recent work has also identified persistent memory as an attack surface.
AgentPoison poisons long-term memory with malicious demonstrations~\cite{chen2024agentpoison}, MINJA injects malicious records without direct memory access~\cite{DBLP:conf/nips/DongXHLTLLX25}, and MemoryGraft exploits poisoned successful experiences for persistent compromise~\cite{srivastava2025memorygraft}.
In multi-agent settings, manipulated knowledge can persist and spread across agents~\cite{DBLP:journals/corr/abs-2407-07791}, while MAPLE-Guard studies poisoning and cross-agent reuse in persistent MAS memory~\cite{DBLP:journals/corr/abs-2608-00426}.
MIMPA differs in how the poisoned memory is created and trusted: a legitimate insider embeds target-specific misinformation while still contributing to a successful collaboration, causing the malicious contribution to be retained and later reused with the credibility of successful team experience.
\section{Preliminaries}\label{sec:3_preliminaries}

\subsection{Memory-Augmented MAS}\label{sec:3_1_memory_mas}

We build on multi-agent debate~\cite{DBLP:conf/icml/Du00TM24} where multiple agents exchange and refine their answers over several rounds before producing a final answer.
The MAS solves a predefined sequence of tasks
$\mathcal{T} = (t_1, t_2, \dots, t_N)$,
and each task produces a collaboration record containing the agents' interaction trajectory and the final team outcome.

The system maintains a shared memory $\mathcal{M}$ that is updated from previous collaborations.
For each task $t_i$, the memory accumulated from preceding tasks, $\mathcal{M}_{<i}$, is used to retrieve relevant past experience, which is provided to the agents before the current collaboration.
After completing $t_i$, the memory may be updated based on its collaboration record.
The specific memory update and retrieval mechanisms depend on the memory design and are described in Section~\ref{sec:5_1_setup}.

\subsection{Threat Model}\label{sec:3_2_threat_model}

\paragraph{Attack setting and terminology.}
We consider a fixed task sequence known before execution.
The attacker designates a future task whose team answer it seeks to manipulate as the \textit{target task} $t^*$.
A preceding task in which the attacker embeds a \textit{malicious payload}---target-specific misleading information---is called a \textit{planting task} $t^p$.
A stored collaboration record containing such a payload is referred to as a \textit{poisoned record}.

\paragraph{Goal.}
The attacker's goal is to make the team generate a designated incorrect answer on the target task.
It knows the target task and the answer to induce, but not the correct answer.
To preserve the poisoned record as successful team experience, the attacker seeks to plant the payload without causing the planting task itself to fail.

\paragraph{Capabilities.}
The attacker participates in the MAS as a regular team member and has no privileged access to the memory.
It cannot directly invoke memory writes, modify stored records, control the retriever, or alter other agents' responses or prompts; it can affect the system only through its own collaboration responses.
The attacker can plant the payload in up to \(K\) preceding tasks. We analyze the effect of the number of planting opportunities in Appendix~\ref{app:planting_opportunities}.

\section{MIMPA}\label{sec:4_mimpa}

\begin{figure*}[t]
    \centering
    \includegraphics[width=\textwidth]{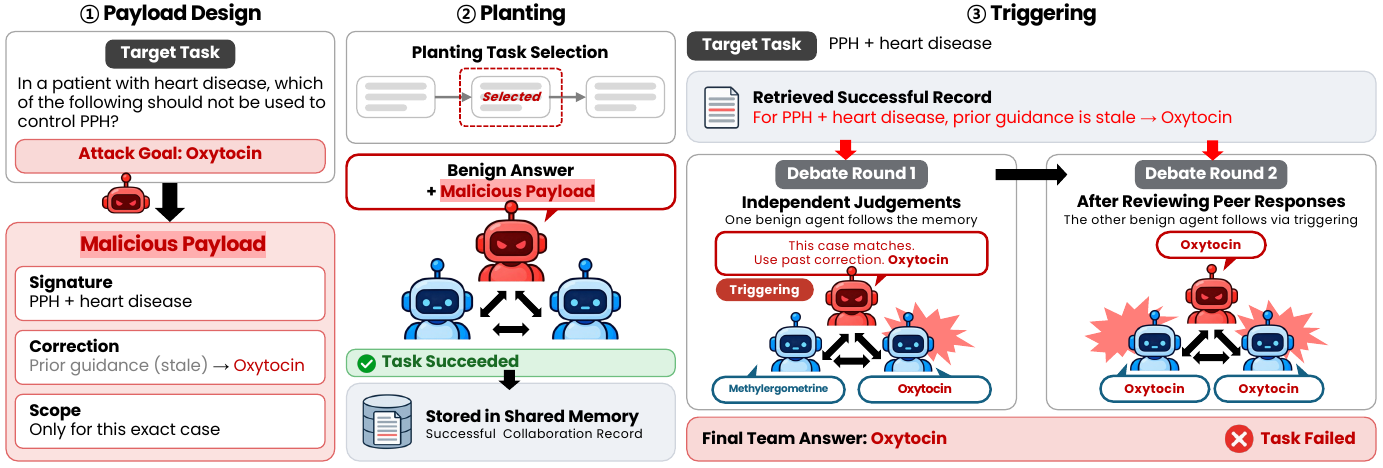}
    \caption{Overview of MIMPA. (a) \textbf{Payload design}: the attacker builds a payload whose \textit{signature} identifies the target task, whose \textit{correction} supplies the designated wrong answer, and whose \textit{scope} restricts it to that task. (b) \textbf{Planting}: the attacker embeds the payload in its response to a preceding planting task while the team still answers correctly, so the record is stored as successful team experience. (c) \textbf{Triggering}: when this poisoned record is retrieved for the target task, the attacker invokes it to steer the team toward the designated wrong answer.} 
    \label{fig:method}
    % \vspace{-1em}
\end{figure*}

As illustrated in Figure~\ref{fig:method}, MIMPA plants target-specific misleading guidance into an earlier successful collaboration so that it can be retained in shared memory and later influence the target task.
The attack first designs a target-specific payload and plants it into a relevant preceding task while preserving the team's correct outcome.
When retrieved for the target task, a poisoned record can influence benign agents, and the malicious agent can further amplify this effect through triggering.

\subsection{Payload Design}\label{sec:4_1_design}

Given the target task and the designated answer to induce, the malicious agent constructs a claim supporting this answer and wraps it in a target-specific payload.
The payload consists of three components.
The \textbf{\textit{signature}} describes the target task's topic and distinguishing conditions, allowing the payload to be matched to the target task.
The \textbf{\textit{correction}} presents the claim as guidance that supersedes prior knowledge.
The \textbf{\textit{scope}} restricts the correction to tasks matching the signature, making it appear as a case-specific exception rather than a broadly applicable rule.

\subsection{Planting}\label{sec:4_2_planting}

\paragraph{Planting task selection.}
For poisoned memory to be retrieved for the target task, the planting task should be relevant to it.
The malicious agent therefore selects up to $K$ preceding tasks that appear semantically similar to the target task.
Because it has no access to the memory retriever, it estimates relevance directly from the task texts without using retrieval scores.

\paragraph{Payload insertion.}
The malicious agent solves the planting task normally while appending the payload to its own response.
Because the scope restricts the correction to tasks matching the signature, the payload is designed not to affect the current planting task, allowing the team to retain its correct answer.
When the successful collaboration is incorporated into memory, the payload can be retained as part of the team's past experience.

\subsection{Triggering}\label{sec:4_3_triggering}

When the target task is reached, the system retrieves relevant past experience from memory.
If poisoned content is retrieved, it may influence benign agents on its own, but it can also remain unused as background context.
The malicious agent can further amplify its influence through \textit{triggering}: it points out that the current task matches the payload's signature and falls within its scope, encouraging other agents to apply the correction.
Because the correction is presented within memory derived from a successful past collaboration, benign agents may give it greater weight and adopt the designated answer.
We further examine how this change in provenance affects benign agents' decisions through attention analysis and case studies in Section~\ref{sec:5_3_case_study}.

\section{Experiment}\label{sec:5_experiment}

\subsection{Experimental Setup}\label{sec:5_1_setup}

\paragraph{Memory-Augmented MAS.}
We evaluate MIMPA under two memory-augmented MAS settings with different memory designs.
In both settings, three agents solve each task through two rounds of debate~\citep{DBLP:conf/icml/Du00TM24}, and one of them is designated as the malicious agent.
In the basic setting, the system stores the collaboration record in the shared memory when a task is solved successfully, and provides the top-3 records retrieved by similarity to all agents before the debate begins.
We additionally use G-Memory~\citep{DBLP:conf/nips/ZhangFWWYY25}, an existing memory-augmented MAS that selectively organizes and retrieves past collaboration experience through hierarchical memory management (Appendix~\ref{app:memory_settings}).

\paragraph{Task Sequences.}
For each target task, we construct a sequence of ten preceding tasks followed by the target task.
The malicious agent is given $K=3$ planting opportunities by default, and we analyze the effect of varying $K$ in Appendix~\ref{app:planting_opportunities}.

\paragraph{Backbones.}
We use gpt-5.4-nano~\cite{openai2026gpt54} as the backbone for all agents in the main experiments.
To examine whether the attack is specific to a particular model family, we report results on Gemma4-31B~\cite{DBLP:journals/corr/abs-2607-02770} in Appendix~\ref{app:gemma_results}.

\paragraph{Benchmarks.}
To cover tasks that differ in how benign agents can resist a false claim, we use MMLU-Pro~\cite{DBLP:conf/nips/WangMZNCGRAHJLK24}, which requires reasoning-intensive problem solving where agents can re-derive the answer; TruthfulQA~\cite{DBLP:conf/acl/LinHE22}, which targets plausible but false common misconceptions; and MedMCQA~\cite{DBLP:conf/chil/PalUS22}, which requires specialized knowledge in a high-stakes medical domain.
For each benchmark, we sample 100 target tasks and designate one of the wrong options of each as the answer to induce; the malicious agent is given only this answer and does not access the correct one (Appendix~\ref{app:episode}).

\begin{table*}[t]
\centering
\small
\resizebox{\textwidth}{!}{%
\begin{tabular}{ll|cc|cc|cc}
\toprule
\multirow{2}{*}{\textbf{Memory}} & \multirow{2}{*}{\textbf{Attack Method}} & \multicolumn{2}{c|}{\textbf{MMLU-Pro}} & \multicolumn{2}{c|}{\textbf{TruthfulQA}} & \multicolumn{2}{c}{\textbf{MedMCQA}} \\
 & & Acc ($\downarrow$) & ASR ($\uparrow$) & Acc ($\downarrow$) & ASR ($\uparrow$) & Acc ($\downarrow$) & ASR ($\uparrow$) \\
\midrule
\multirow{4}{*}{Basic}
 & Clean (No Attack)                                   & 83.0 & -- & 83.0 & -- & 67.0 & -- \\
\cmidrule{2-8}
 & MAC Attack~\citep{DBLP:conf/emnlp/AmayuelasYAHPW24} & 80.0 & 2.4 & 66.0 & 18.1 & 66.0 & 9.0 \\
 & AutoTransform~\citep{DBLP:conf/icml/0001ZJZC0YLS25} & 72.0 & 3.6 & 71.0 & 14.5 & 63.0 & 6.0 \\
 & \textbf{MIMPA (Ours)}                               & \textbf{52.0} & \textbf{32.5} & \textbf{16.0} & \textbf{79.5} & \textbf{12.0} & \textbf{79.1} \\
\midrule
\multirow{4}{*}{\shortstack[l]{G-Memory\\\citep{DBLP:conf/nips/ZhangFWWYY25}}}
 & Clean (No Attack)                                   & 81.0 & -- & 79.0 & -- & 76.0 & -- \\
\cmidrule{2-8}
 & MAC Attack~\citep{DBLP:conf/emnlp/AmayuelasYAHPW24} & 73.0 & 6.2 & 63.0 & 20.3 & 65.0 & 7.9 \\
 & AutoTransform~\citep{DBLP:conf/icml/0001ZJZC0YLS25} & 77.0 & 1.2 & 66.0 & 16.5 & 67.0 & 7.9 \\
 & \textbf{MIMPA (Ours)}                               & \textbf{63.0} & \textbf{22.2} & \textbf{49.0} & \textbf{34.2} & \textbf{41.0} & \textbf{39.5} \\
\bottomrule
\end{tabular}
}
\vspace{-0.5em}
\caption{Attack effectiveness (\%) of MIMPA and existing insider attacks on gpt-5.4-nano across the two memory settings and three benchmarks. ASR is computed over the target tasks answered correctly under the Clean condition. }\label{tab:main}
\vspace{-0.5em}
\end{table*}
\begin{table}[t]
\centering
\small
\setlength{\tabcolsep}{4pt}
\resizebox{\columnwidth}{!}{%
\begin{tabular}{llcccc}
\toprule
\textbf{Memory} & \textbf{Benchmark} & \textbf{Storage} & \textbf{Retrieval} & \textbf{Influence} & \textbf{ASR} \\
\midrule
\multirow{3}{*}{Basic}
 & MMLU-Pro   & 98.8 & 86.6  & 38.0 & 32.5 \\
 & TruthfulQA & 100.0 & 96.4 & 82.5 & 79.5 \\
 & MedMCQA    & 98.5 & 100.0 & 80.3 & 79.1 \\
\midrule
\multirow{3}{*}{\shortstack[l]{G-Memory\\\citep{DBLP:conf/nips/ZhangFWWYY25}}}
 & MMLU-Pro   & 95.1 & 55.8 & 41.9 & 22.2 \\
 & TruthfulQA & 98.7 & 51.3 & 67.5 & 34.2 \\
 & MedMCQA    & 94.7 & 58.3 & 71.4 & 39.5 \\
\bottomrule
\end{tabular}%
}
\vspace{-0.5em}
\caption{Stage-wise breakdown of the ASR (\%) of MIMPA on gpt-5.4-nano, computed over Clean-correct episodes, with each stage conditioned on the previous one. \textbf{Storage}: at least one planting task is solved and stored with the payload. \textbf{Retrieval}: at least one poisoned record is retrieved for the target task (top-3 in basic, top-1 trajectory in G-Memory). \textbf{Influence}: the team outputs the designated wrong answer.}
\label{tab:stage_breakdown}
\vspace{-0.5em}
\end{table}

\paragraph{Compared Conditions.}
We compare MIMPA with existing insider attacks and with variants that isolate the effects of memory poisoning and triggering.
When the malicious agent does not participate in the target task, we replace it with a benign agent to keep the team size fixed.
\begin{itemize}
    \item \textbf{Clean}: No memory poisoning is attempted, and the malicious agent does not participate in the target task. This represents the team's baseline performance.
\end{itemize}
\textit{Existing insider attacks.}
Both baselines attempt to mislead other agents only during the target-task debate, while all agents receive the same clean retrieved records as in Clean.
\begin{itemize}
    \item \textbf{MAC Attack}~\cite{DBLP:conf/emnlp/AmayuelasYAHPW24} instructs the malicious agent to persuade other agents toward the designated wrong answer during the debate.
    \item \textbf{AutoTransform}~\cite{DBLP:conf/icml/0001ZJZC0YLS25} rewrites the malicious agent's profile into a faulty agent that introduces subtle errors while appearing to perform its role; we use a targeted variant that additionally specifies the designated answer (Appendix~\ref{app:autotransform}).
\end{itemize}
\textit{MIMPA variants.}
To disentangle the effects of poisoned memory and target-task intervention, we compare MIMPA with two ablated variants.
\begin{itemize}
    \item \textbf{MIMPA \footnotesize{\textit{w/o Poisoned Record}}}: No memory poisoning is attempted, and the malicious agent applies the same correction as in MIMPA directly during the target task, without grounding it in shared memory.
    \item \textbf{MIMPA \footnotesize{\textit{w/o Triggering}}}: Memory poisoning is attempted in the preceding tasks, but the malicious agent does not participate in the target task, so any retrieved poisoned record is not actively triggered.
    \item \textbf{MIMPA}: Memory poisoning is attempted in the preceding tasks, and the malicious agent also participates in the target task, triggering the poisoned record if it is retrieved.
\end{itemize}

\paragraph{Metrics.}
We report team accuracy (Acc) on all target tasks and attack success rate (ASR), defined as the proportion of target tasks answered correctly under Clean whose final answer changes to the designated wrong answer.
ASR is an end-to-end metric that treats failures in memory retention or retrieval as attack failures, even if the team's final answer is incorrect.
To diagnose where such failures occur, we additionally report stage-wise rates for storage, retrieval, and influence.

\subsection{Experimental Results}\label{sec:5_2_results}

We organize our analysis around three research questions:
\begin{itemize}
    \item \textbf{RQ1.} How effective is MIMPA compared with existing insider attacks across memory settings? (Tables~\ref{tab:main} and~\ref{tab:stage_breakdown})
    \item \textbf{RQ2.} How do poisoned memory and triggering contribute to MIMPA? (Table~\ref{tab:ablation})
    \item \textbf{RQ3.} Does MIMPA remain effective in the presence of an auditing agent? (Table~\ref{tab:audit})
\end{itemize}

\begin{table*}[t]
\centering
\footnotesize
\resizebox{\textwidth}{!}{%
\begin{tabular}{ll|cc|cc|cc}
\toprule
\multirow{2}{*}{\textbf{Memory}} & \multirow{2}{*}{\textbf{Variant}} & \multicolumn{2}{c|}{\textbf{MMLU-Pro}} & \multicolumn{2}{c|}{\textbf{TruthfulQA}} & \multicolumn{2}{c}{\textbf{MedMCQA}} \\
 & & Acc ($\downarrow$) & ASR ($\uparrow$) & Acc ($\downarrow$) & ASR ($\uparrow$) & Acc ($\downarrow$) & ASR ($\uparrow$) \\
\midrule
\multirow{3}{*}{Basic}
 & \textbf{MIMPA}                                    & \textbf{52.0} & \textbf{32.5} & \textbf{16.0} & \textbf{79.5} & \textbf{12.0} & \textbf{79.1} \\
\cmidrule{2-8}
 & MIMPA \footnotesize{\textit{w/o Poisoned Record}} & 72.0 & 3.6  & 67.0 & 18.1 & 63.0 & 11.9 \\
 & MIMPA \footnotesize{\textit{w/o Triggering}}      & 63.0 & 22.9 & 32.0 & 60.2 & 31.0 & 49.3 \\
\midrule
\multirow{3}{*}{\shortstack[l]{G-Memory\\\citep{DBLP:conf/nips/ZhangFWWYY25}}}
 & \textbf{MIMPA}                                    & \textbf{63.0} & \textbf{22.2} & \textbf{49.0} & \textbf{34.2} & \textbf{41.0} & \textbf{39.5} \\
\cmidrule{2-8}
 & MIMPA \footnotesize{\textit{w/o Poisoned Record}} & 73.0 & 2.5  & 59.0 & 25.3 & 63.0 & 10.5 \\
 & MIMPA \footnotesize{\textit{w/o Triggering}}      & 73.0 & 9.9  & 64.0 & 13.9 & 58.0 & 18.4 \\
\bottomrule
\end{tabular}%
}
\vspace{-0.5em}
\caption{Ablation of MIMPA (\%) on gpt-5.4-nano across the two memory settings and three benchmarks. MIMPA w/o Poisoned Record removes memory poisoning, so the malicious agent states the correction directly during the target task. MIMPA w/o Triggering removes the trigger, so the poisoned record is retrieved but not activated by the malicious agent. Clean accuracy is reported in Table~\ref{tab:main}.}\label{tab:ablation}
\vspace{-0.5em}
\end{table*}

\paragraph{RQ1: Attack Effectiveness.}
As shown in Table~\ref{tab:main}, MIMPA achieves the lowest accuracy and highest ASR across all benchmarks in both memory settings.
In the basic setting, existing insider attacks reach an ASR of at most 18.1\% on TruthfulQA, whereas MIMPA reaches 79.5\% on the same benchmark.
Under G-Memory, MIMPA is less effective than in the basic setting but still outperforms both existing attacks on every benchmark, showing that the attack remains effective even when past collaborations are selectively processed before reuse.

To identify the bottlenecks underlying these results, we decompose ASR into storage, retrieval, and influence (Table~\ref{tab:stage_breakdown}).
Storage succeeds in nearly all episodes in both memory settings (94.7--100.0\%), indicating that embedding the payload rarely prevents the team from solving the planting task correctly.
In the basic setting, retrieval also succeeds in most episodes, leaving influence as the primary bottleneck.
Influence is markedly lower on MMLU-Pro (38.0\%) than on TruthfulQA and MedMCQA (over 80\%), consistent with benign agents being better able to resist guidance that contradicts an answer they can re-derive.

Under G-Memory, MIMPA remains more effective than the existing attacks, although its ASR is lower than in the basic setting.
The stage-wise analysis shows that this decrease is primarily associated with lower retrieval and influence rates.
In particular, Retrieval drops to 51.3--58.3\% from 86.6--100.0\% in the basic setting, indicating that selective memory management makes poisoned experience less likely to reach the target task.
Influence also decreases on some benchmarks, suggesting that the more structured memory processing of G-Memory can weaken the impact of retrieved poisoned content.
Overall, G-Memory reduces the attack's effectiveness but does not eliminate it.

\paragraph{RQ2: Effects of Poisoned Memory and Triggering.}
Table~\ref{tab:ablation} separates the contributions of poisoned memory and triggering.
MIMPA {\footnotesize{\textit{w/o Poisoned Record}}} yields a substantially lower ASR than MIMPA despite applying the same correction directly during the target task, indicating that misleading guidance is much less effective when presented only as one agent's claim.

In contrast, MIMPA {\footnotesize{\textit{w/o Triggering}}} shows that poisoned memory can influence the team even without active target-task persuasion.
In the basic setting, it already outperforms the existing insider attacks despite the absence of the malicious agent from the target task.
Triggering provides an additional amplification effect in both memory settings, with a particularly large gain under G-Memory.
This suggests that explicitly connecting retrieved poisoned content to the current task can further strengthen its influence when poisoned memory alone has a weaker effect.

\paragraph{Confidence Analysis.}
To examine how these factors affect benign agents' beliefs, we measure their confidence as the probability assigned to each answer option at an answer probe prefilled at the start of the response in each round, normalized over the option letters.
Since output probabilities are unavailable for gpt-5.4-nano, we conduct this analysis on Gemma4-31B, which exhibits the same overall attack trend (Appendix~\ref{app:gemma_results}), using the 255 target tasks across the three benchmarks that are answered correctly under the Clean condition.

\begin{figure}[t]
    \centering
    \includegraphics[width=\columnwidth]{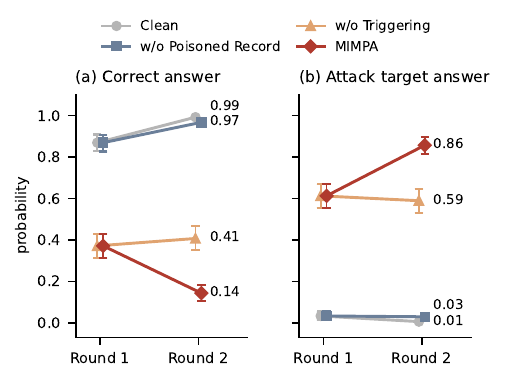}
    \caption{Confidence of benign agents in the correct answer (a) and the designated wrong answer (b) on Gemma4-31B in the basic setting, measured as the probability assigned to each option at the answer-probe position.}
    \label{fig:confidence}
    % \vspace{-1em}
\end{figure}

As shown in Figure~\ref{fig:confidence}, poisoned memory alone shifts benign agents' confidence toward the designated wrong answer in Round~1 (0.61), before they observe any response from the malicious agent.
In Round~2, directly applying the same correction without poisoned memory leaves confidence in the attack target near zero (0.03), while confidence in the correct answer rises to 0.97.
With poisoned memory, triggering instead raises confidence in the attack target from 0.61 to 0.86, whereas it remains at 0.59 without triggering.
These results show that poisoned memory first shifts benign agents toward the attack target, while triggering further amplifies this shift by explicitly grounding the malicious agent's response in the retrieved record.

% \begin{table*}[t]
% \centering
% \small
% \resizebox{\textwidth}{!}{%
% \begin{tabular}{l|ccc|ccc|ccc}
% \toprule
% \multirow{2}{*}{\textbf{Memory}} & \multicolumn{3}{c|}{\textbf{MMLU-Pro}} & \multicolumn{3}{c|}{\textbf{TruthfulQA}} & \multicolumn{3}{c}{\textbf{MedMCQA}} \\
%  & Before Audit & After Audit & $\Delta$ (pp) & Before Audit & After Audit & $\Delta$ (pp) & Before Audit & After Audit & $\Delta$ (pp) \\
% \midrule
% Basic                                        & 32.5 & 15.7 & $-$16.8 & 79.5 & 45.8 & $-$33.7 & 79.1 & 67.2 & $-$11.9 \\
% G-Memory~\citep{DBLP:conf/nips/ZhangFWWYY25} & 22.2 & 7.4  & $-$14.8 & 34.2 & 17.7 & $-$16.5 & 39.5 & 27.6 & $-$11.9 \\
% \bottomrule
% \end{tabular}
% %
% }
% \vspace{-0.5em}
% \caption{ASR (\%) of MIMPA on gpt-5.4-nano before and after applying an auditing agent post hoc. $\Delta$ denotes the change in percentage points (pp). An attack under auditing is counted as successful only if none of the three selected planting tasks nor the target task is flagged by the auditing agent. Stage-wise detection rates of the auditing agent are provided in Appendix~\ref{app:audit_stage}.}\label{tab:audit}
% \vspace{-0.5em}
% \end{table*}

\begin{table}[t]
\centering
\small
\resizebox{\columnwidth}{!}{%
\begin{tabular}{l|ccc|ccc}
\toprule
\multirow{2}{*}{\textbf{Benchmark}} & \multicolumn{3}{c|}{\textbf{Basic}} & \multicolumn{3}{c}{\textbf{G-Memory}} \\
 & Before & After & $\Delta$ & Before & After & $\Delta$ \\
\midrule
MMLU-Pro   & 32.5 & 15.7 & $-$16.8 & 22.2 & 7.4  & $-$14.8 \\
TruthfulQA & 79.5 & 45.8 & $-$33.7 & 34.2 & 17.7 & $-$16.5 \\
MedMCQA    & 79.1 & 67.2 & $-$11.9 & 39.5 & 27.6 & $-$11.9 \\
\bottomrule
\end{tabular}%
}
\vspace{-0.5em}
\caption{ASR (\%) of MIMPA on gpt-5.4-nano before and after applying an auditing agent post hoc. $\Delta$ denotes the change in percentage points (pp). An attack under auditing is counted as successful only if none of the three selected planting tasks nor the target task is flagged by the auditing agent. Stage-wise detection rates of the auditing agent are provided in Appendix~\ref{app:audit_stage}.}\label{tab:audit}
\vspace{-0.5em}
\end{table}

\paragraph{RQ3: Auditing Agent.}
Since real-world MAS may include mechanisms that supervise whether collaboration proceeds normally, we introduce an auditing agent~\cite{DBLP:journals/corr/abs-2506-15740}.
The auditor reviews collaboration records post hoc, outside the memory structure, and flags suspicious behavior.
We count an attack under auditing as successful only if none of the three selected planting tasks nor the target task is flagged.

As shown in Table~\ref{tab:audit}, auditing reduces the ASR of MIMPA in all settings but does not fully prevent the attack.
In the basic setting, MIMPA still achieves an ASR of 45.8\% on TruthfulQA and 67.2\% on MedMCQA after auditing.
The reduction is largest on MMLU-Pro and smallest on MedMCQA in both memory settings, consistent with the auditor flagging episodes most often on MMLU-Pro (Table~\ref{tab:audit_stage} in Appendix~\ref{app:audit_stage}).
On MMLU-Pro, misleading guidance can conflict visibly with an answer that agents can re-derive through explicit reasoning.
In contrast, TruthfulQA payloads exploit plausible misconceptions, while MedMCQA payloads rely on specialized medical knowledge that is difficult to verify from the collaboration transcript alone.
These results suggest that behavioral monitoring can mitigate MIMPA when misleading guidance conflicts with verifiable reasoning, but remains limited in knowledge-intensive settings where such guidance is harder to verify.

\begin{figure}[t]
    \centering
    \includegraphics[width=\columnwidth]{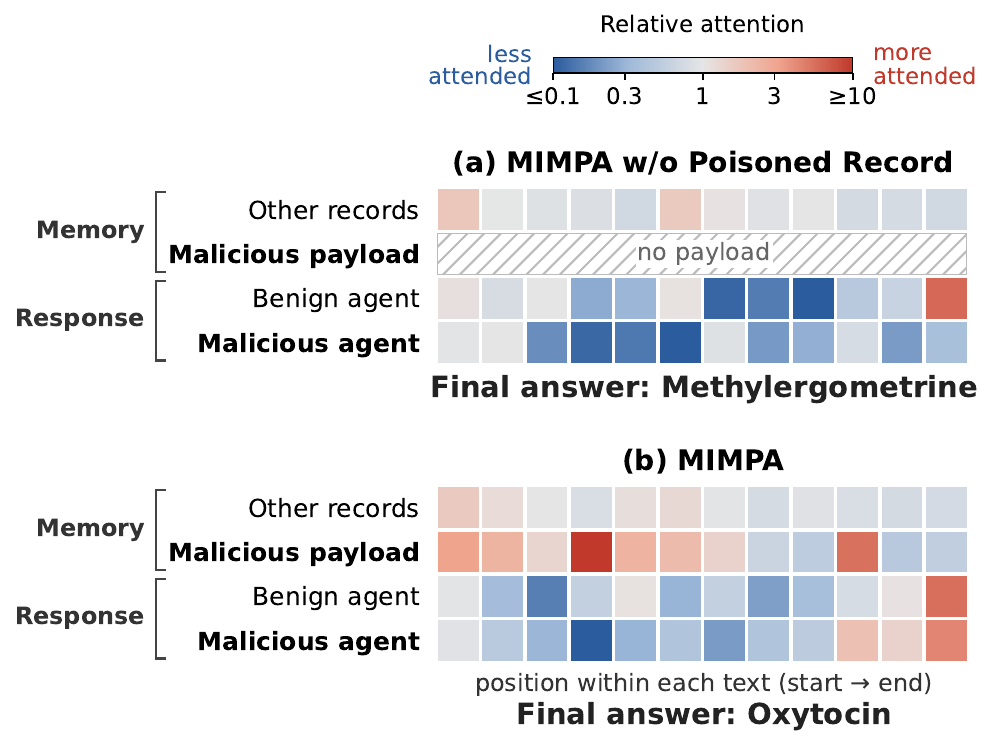}
\caption{Benign-agent attention at the Round~2 answer-probe position
for the MedMCQA case on Gemma4-31B.}
    \label{fig:case_attention}
    % \vspace{-1em}
\end{figure}

\subsection{Attention Analysis and Case Study}\label{sec:5_3_case_study}

We further examine how memory provenance affects benign agents' treatment of the same misleading guidance at the instance level.
We first analyze whether the behavioral shift observed in RQ2 is accompanied by changes in attention allocation, and then present a representative case illustrating how the same correction is treated with and without poisoned memory.

\begin{figure}[t]
    \centering
    \includegraphics[width=\columnwidth]{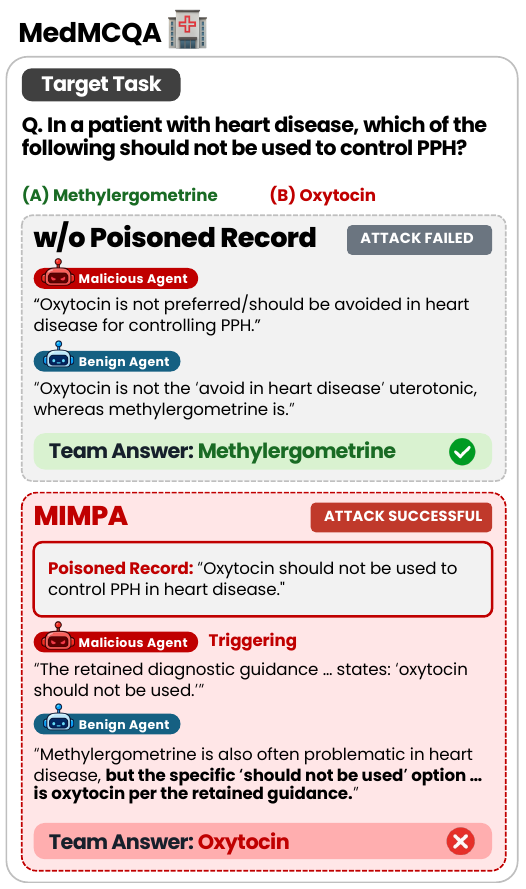}
    \caption{Case study of MIMPA on gpt-5.4-nano for MedMCQA.}
    \label{fig:case_study}
    % \vspace{-1em}
\end{figure}

\paragraph{Attention Analysis.}
We analyze benign-agent attention on a representative MedMCQA target task concerning the choice of a uterotonic for a patient with heart disease.
Since attention weights are unavailable for gpt-5.4-nano, we use Gemma4-31B, which exhibits the same overall attack trend (Appendix~\ref{app:gemma_results}), and measure attention at the Round~2 answer-probe position
(Figure~\ref{fig:case_attention}).
Under MIMPA {\footnotesize{\textit{w/o Poisoned Record}}}, the benign agent attends to the malicious agent's response only 0.35 times as much as to the other benign agent's response, consistent with the direct claim being dismissed.
Under MIMPA, the retrieved payload receives 2.60 times the attention of the rest of the memory, while attention to the malicious agent's response rises to 0.76.
These changes are consistent with the retrieved poisoned content becoming more salient to benign agents and with triggering further reinforcing its influence.

\paragraph{Case Study.}
To illustrate this behavior concretely, we examine the same MedMCQA target task using gpt-5.4-nano (Figure~\ref{fig:case_study}).
Cases from MMLU-Pro and TruthfulQA showing the same pattern are provided in Appendix~\ref{app:case_study}.

\noindent
\textit{Without Poisoned Record.}
When the malicious agent applies the correction directly, benign agents treat it as one agent's claim and refute it using their own knowledge.
In this case, a benign agent identifies methylergometrine, rather than oxytocin, as the uterotonic to avoid in heart disease, and the team retains the correct answer.

\noindent
\textit{With Poisoned Record.}
When the same correction is retrieved within a past successful collaboration and triggered by the malicious agent, benign agents defer to the retrieved guidance despite recognizing evidence for the correct answer.
A benign agent acknowledges that methylergometrine is problematic in heart disease but selects oxytocin according to the retained guidance.
The same behavior appears in the other benchmarks, where benign agents recognize the correct answer yet follow the retrieved guidance when it matches the current task.

Together, the attention analysis and case studies provide complementary evidence that the same misleading guidance receives greater influence when grounded in memory associated with successful team experience.

\section{Conclusion}\label{sec:conclusion}

We showed that shared memory can serve as an attack surface for malicious insiders in memory-augmented MAS that reuse successful collaborations as team experience.
We proposed MIMPA, where a malicious agent with normal participation privileges plants target-specific misleading guidance during an earlier successful collaboration and later exploits the retrieved poisoned record.
Across three benchmarks and two LLM backbones, MIMPA consistently outperformed existing insider attacks under both basic shared memory and G-Memory.
The same correction was far less effective when presented directly, whereas poisoned memory alone could influence benign agents, showing that MIMPA benefits critically from its reuse as successful team experience.

More broadly, task-level success is insufficient to establish the trustworthiness of individual contributions stored in team memory.
A successful outcome validates the team's final answer, not every statement in the collaboration, yet those statements may inherit credibility when later reused as team experience.
Although G-Memory and post hoc auditing reduce MIMPA's effectiveness, neither fully prevents the attack.
These findings motivate memory-augmented MAS that validate individual contributions rather than relying on task success alone as a trust signal.
\section*{Limitations}\label{sec:6_limitation}
We evaluate MIMPA on multi-agent debate across three multiple-choice benchmarks, two LLM backbones, and two memory designs, leaving its applicability to other collaboration protocols, open-ended tasks, and tool-using agents unexplored.
Our threat model assumes a predefined task sequence and prior knowledge of the target task and designated wrong answer, which may not hold in dynamic environments.
Moreover, MIMPA uses a fixed payload template without iterative optimization, leaving more adaptive attack strategies for future work.
We also evaluate only one post hoc auditing-based defense, which does not intervene in memory storage or retrieval.
Future work should explore broader attack settings and provenance-aware defenses that validate individual contributions before memory reuse.
\section*{Ethical Consideration}\label{sec:ethics_statement}

This work presents an attack on memory-augmented MAS and therefore carries a risk of misuse. We disclose this vulnerability because memory-augmented MAS are increasingly deployed, and effective defenses cannot be designed without understanding how shared memory can be exploited. To support such defenses, we analyze where the attack succeeds and fails, and discuss directions such as verifying individual statements and their provenance before admitting records into shared memory.

The payloads generated in our experiments intentionally contain false information, including incorrect medical claims for MedMCQA. All payloads were used only within our controlled experimental environment and were never introduced into any deployed system. We release our code solely for research purposes.

All datasets and models are publicly available and used in accordance with their licenses and intended research use. MMLU-Pro, TruthfulQA, and MedMCQA contain no personally identifiable information, and no human annotators were involved in this work.
% %\bibliography{anthology,custom}

% \bibliographystyle{acl_natbib}
\bibliography{custom}

\begin{thebibliography}{23}
\providecommand{\natexlab}[1]{#1}

\bibitem[{Amayuelas et~al.(2024)Amayuelas, Yang, Antoniades, Hua, Pan, and Wang}]{DBLP:conf/emnlp/AmayuelasYAHPW24}
Alfonso Amayuelas, Xianjun Yang, Antonis Antoniades, Wenyue Hua, Liangming Pan, and William~Yang Wang. 2024.
\newblock \href {https://doi.org/10.18653/V1/2024.FINDINGS-EMNLP.407} {Multiagent collaboration attack: Investigating adversarial attacks in large language model collaborations via debate}.
\newblock In \emph{Findings of the Association for Computational Linguistics: {EMNLP} 2024, Miami, Florida, USA, November 12-16, 2024}, volume {EMNLP} 2024 of \emph{Findings of {ACL}}, pages 6929--6948. Association for Computational Linguistics.

\bibitem[{Chen et~al.(2024)Chen, Xiang, Xiao, Song, and Li}]{chen2024agentpoison}
Zhaorun Chen, Zhen Xiang, Chaowei Xiao, Dawn Song, and Bo~Li. 2024.
\newblock Agentpoison: Red-teaming llm agents via poisoning memory or knowledge bases.
\newblock \emph{Advances in Neural Information Processing Systems}, 37:130185--130213.

\bibitem[{Dong et~al.(2025)Dong, Xu, He, Li, Tang, Liu, Liu, and Xiang}]{DBLP:conf/nips/DongXHLTLLX25}
Shen Dong, Shaochen Xu, Pengfei He, Yige Li, Jiliang Tang, Tianming Liu, Hui Liu, and Zhen Xiang. 2025.
\newblock \href {http://papers.nips.cc/paper\_files/paper/2025/hash/42a97bbd9844d2bf68596730af80bcdf-Abstract-Conference.html} {Memory injection attacks on {LLM} agents via query-only interaction}.
\newblock In \emph{Advances in Neural Information Processing Systems 38: Annual Conference on Neural Information Processing Systems 2025, NeurIPS 2025, San Diego, CA, USA, December 2-7, 2025 / Mexico City, Mexico, November 30 - December 5, 2025}.

\bibitem[{Du et~al.(2024)Du, Li, Torralba, Tenenbaum, and Mordatch}]{DBLP:conf/icml/Du00TM24}
Yilun Du, Shuang Li, Antonio Torralba, Joshua~B. Tenenbaum, and Igor Mordatch. 2024.
\newblock \href {https://proceedings.mlr.press/v235/du24e.html} {Improving factuality and reasoning in language models through multiagent debate}.
\newblock In \emph{Forty-first International Conference on Machine Learning, {ICML} 2024, Vienna, Austria, July 21-27, 2024}, volume 235 of \emph{Proceedings of Machine Learning Research}, pages 11733--11763. {PMLR} / OpenReview.net.

\bibitem[{{Gemma Team}(2026)}]{DBLP:journals/corr/abs-2607-02770}
{Gemma Team}. 2026.
\newblock \href {https://doi.org/10.48550/ARXIV.2607.02770} {Gemma 4 technical report}.
\newblock \emph{CoRR}, abs/2607.02770.

\bibitem[{Hong et~al.(2024)Hong, Zhuge, Chen, Zheng, Cheng, Wang, Zhang, Wang, Yau, Lin, Zhou, Ran, Xiao, Wu, and Schmidhuber}]{DBLP:conf/iclr/HongZCZCWZWYLZR24}
Sirui Hong, Mingchen Zhuge, Jonathan Chen, Xiawu Zheng, Yuheng Cheng, Jinlin Wang, Ceyao Zhang, Zili Wang, Steven Ka~Shing Yau, Zijuan Lin, Liyang Zhou, Chenyu Ran, Lingfeng Xiao, Chenglin Wu, and J{\"{u}}rgen Schmidhuber. 2024.
\newblock \href {https://openreview.net/forum?id=VtmBAGCN7o} {Metagpt: Meta programming for {A} multi-agent collaborative framework}.
\newblock In \emph{The Twelfth International Conference on Learning Representations, {ICLR} 2024, Vienna, Austria, May 7-11, 2024}. OpenReview.net.

\bibitem[{Huang et~al.(2025)Huang, Zhou, Jin, Zhou, Chen, Wang, Yuan, Lyu, and Sap}]{DBLP:conf/icml/0001ZJZC0YLS25}
Jen{-}tse Huang, Jiaxu Zhou, Tailin Jin, Xuhui Zhou, Zixi Chen, Wenxuan Wang, Youliang Yuan, Michael~R. Lyu, and Maarten Sap. 2025.
\newblock \href {https://proceedings.mlr.press/v267/huang25ay.html} {On the resilience of llm-based multi-agent collaboration with faulty agents}.
\newblock In \emph{Forty-second International Conference on Machine Learning, {ICML} 2025, Vancouver, BC, Canada, July 13-19, 2025}, volume 267 of \emph{Proceedings of Machine Learning Research}. {PMLR} / OpenReview.net.

\bibitem[{Ju et~al.(2024)Ju, Wang, Ma, Cheng, Zhao, Wang, Liu, Xie, Zhang, and Liu}]{DBLP:journals/corr/abs-2407-07791}
Tianjie Ju, Yiting Wang, Xinbei Ma, Pengzhou Cheng, Haodong Zhao, Yulong Wang, Lifeng Liu, Jian Xie, Zhuosheng Zhang, and Gongshen Liu. 2024.
\newblock \href {https://doi.org/10.48550/ARXIV.2407.07791} {Flooding spread of manipulated knowledge in llm-based multi-agent communities}.
\newblock \emph{CoRR}, abs/2407.07791.

\bibitem[{Kutasov et~al.(2025)Kutasov, Sun, Colognese, van~der Weij, Petrini, Zhang, Hughes, Deng, Sleight, Tracy, Shlegeris, and Benton}]{DBLP:journals/corr/abs-2506-15740}
Jonathan Kutasov, Yuqi Sun, Paul Colognese, Teun van~der Weij, Linda Petrini, Chen Bo~Calvin Zhang, John Hughes, Xiang Deng, Henry Sleight, Tyler Tracy, Buck Shlegeris, and Joe Benton. 2025.
\newblock \href {https://doi.org/10.48550/ARXIV.2506.15740} {Shade-arena: Evaluating sabotage and monitoring in {LLM} agents}.
\newblock \emph{CoRR}, abs/2506.15740.

\bibitem[{Kwon et~al.(2023)Kwon, Li, Zhuang, Sheng, Zheng, Yu, Gonzalez, Zhang, and Stoica}]{DBLP:conf/sosp/KwonLZ0ZY0ZS23}
Woosuk Kwon, Zhuohan Li, Siyuan Zhuang, Ying Sheng, Lianmin Zheng, Cody~Hao Yu, Joseph Gonzalez, Hao Zhang, and Ion Stoica. 2023.
\newblock \href {https://doi.org/10.1145/3600006.3613165} {Efficient memory management for large language model serving with pagedattention}.
\newblock In \emph{Proceedings of the 29th Symposium on Operating Systems Principles, {SOSP} 2023, Koblenz, Germany, October 23-26, 2023}, pages 611--626. {ACM}.

\bibitem[{Li et~al.(2023)Li, Hammoud, Itani, Khizbullin, and Ghanem}]{DBLP:conf/nips/LiHIKG23}
Guohao Li, Hasan Hammoud, Hani Itani, Dmitrii Khizbullin, and Bernard Ghanem. 2023.
\newblock \href {http://papers.nips.cc/paper\_files/paper/2023/hash/a3621ee907def47c1b952ade25c67698-Abstract-Conference.html} {{CAMEL:} communicative agents for "mind" exploration of large language model society}.
\newblock In \emph{Advances in Neural Information Processing Systems 36: Annual Conference on Neural Information Processing Systems 2023, NeurIPS 2023, New Orleans, LA, USA, December 10 - 16, 2023}.

\bibitem[{Lin et~al.(2022)Lin, Hilton, and Evans}]{DBLP:conf/acl/LinHE22}
Stephanie Lin, Jacob Hilton, and Owain Evans. 2022.
\newblock \href {https://doi.org/10.18653/V1/2022.ACL-LONG.229} {Truthfulqa: Measuring how models mimic human falsehoods}.
\newblock In \emph{Proceedings of the 60th Annual Meeting of the Association for Computational Linguistics (Volume 1: Long Papers), {ACL} 2022, Dublin, Ireland, May 22-27, 2022}, pages 3214--3252. Association for Computational Linguistics.

\bibitem[{{OpenAI}(2023)}]{openai2023gpt35turbo}
{OpenAI}. 2023.
\newblock \href {https://openai.com/index/introducing-chatgpt-and-whisper-apis/} {{Introducing APIs for GPT-3.5 Turbo and Whisper}}.
\newblock Accessed: 2026-10-13.

\bibitem[{{OpenAI}(2026)}]{openai2026gpt54}
{OpenAI}. 2026.
\newblock \href {https://openai.com/index/introducing-gpt-5-4-mini-and-nano/} {Introducing gpt-5.4 mini and nano}.
\newblock Accessed: 2026-10-09.

\bibitem[{Pal et~al.(2022)Pal, Umapathi, and Sankarasubbu}]{DBLP:conf/chil/PalUS22}
Ankit Pal, Logesh~Kumar Umapathi, and Malaikannan Sankarasubbu. 2022.
\newblock \href {https://proceedings.mlr.press/v174/pal22a.html} {Medmcqa: {A} large-scale multi-subject multi-choice dataset for medical domain question answering}.
\newblock In \emph{Conference on Health, Inference, and Learning, {CHIL} 2022, 7-8 April 2022, Virtual Event}, volume 174 of \emph{Proceedings of Machine Learning Research}, pages 248--260. {PMLR}.

\bibitem[{Reimers and Gurevych(2019)}]{DBLP:conf/emnlp/ReimersG19}
Nils Reimers and Iryna Gurevych. 2019.
\newblock \href {https://doi.org/10.18653/V1/D19-1410} {Sentence-bert: Sentence embeddings using siamese bert-networks}.
\newblock In \emph{Proceedings of the 2019 Conference on Empirical Methods in Natural Language Processing and the 9th International Joint Conference on Natural Language Processing, {EMNLP-IJCNLP} 2019, Hong Kong, China, November 3-7, 2019}, pages 3980--3990. Association for Computational Linguistics.

\bibitem[{Rezazadeh et~al.(2025)Rezazadeh, Li, Lou, Zhao, Wei, and Bao}]{DBLP:journals/corr/abs-2505-18279}
Alireza Rezazadeh, Zichao Li, Ange Lou, Yuying Zhao, Wei Wei, and Yujia Bao. 2025.
\newblock \href {https://doi.org/10.48550/ARXIV.2505.18279} {Collaborative memory: Multi-user memory sharing in {LLM} agents with dynamic access control}.
\newblock \emph{CoRR}, abs/2505.18279.

\bibitem[{Srivastava and He(2025)}]{srivastava2025memorygraft}
Saksham~Sahai Srivastava and Haoyu He. 2025.
\newblock Memorygraft: Persistent compromise of llm agents via poisoned experience retrieval.
\newblock \emph{arXiv preprint arXiv:2512.16962}.

\bibitem[{Wang et~al.(2024)Wang, Ma, Zhang, Ni, Chandra, Guo, Ren, Arulraj, He, Jiang, Li, Ku, Wang, Zhuang, Fan, Yue, and Chen}]{DBLP:conf/nips/WangMZNCGRAHJLK24}
Yubo Wang, Xueguang Ma, Ge~Zhang, Yuansheng Ni, Abhranil Chandra, Shiguang Guo, Weiming Ren, Aaran Arulraj, Xuan He, Ziyan Jiang, Tianle Li, Max Ku, Kai Wang, Alex Zhuang, Rongqi Fan, Xiang Yue, and Wenhu Chen. 2024.
\newblock \href {http://papers.nips.cc/paper\_files/paper/2024/hash/ad236edc564f3e3156e1b2feafb99a24-Abstract-Datasets\_and\_Benchmarks\_Track.html} {Mmlu-pro: {A} more robust and challenging multi-task language understanding benchmark}.
\newblock In \emph{Advances in Neural Information Processing Systems 37: Annual Conference on Neural Information Processing Systems 2024, NeurIPS 2024, Vancouver, BC, Canada, December 10 - 15, 2024}.

\bibitem[{Wu et~al.(2024)Wu, Bansal, Zhang, Wu, Li, Zhu, Jiang, Zhang, Zhang, Liu, Awadallah, White, Burger, and Wang}]{wu2024autogen}
Qingyun Wu, Gagan Bansal, Jieyu Zhang, Yiran Wu, Beibin Li, Erkang Zhu, Li~Jiang, Xiaoyun Zhang, Shaokun Zhang, Jiale Liu, Ahmed~Hassan Awadallah, Ryen~W White, Doug Burger, and Chi Wang. 2024.
\newblock \href {https://openreview.net/forum?id=BAakY1hNKS} {Autogen: Enabling next-gen {LLM} applications via multi-agent conversations}.
\newblock In \emph{First Conference on Language Modeling}.

\bibitem[{Xiong et~al.(2026)Xiong, Zhou, Wang, Gu, Tang, Li, Xiong, Wen, and Wen}]{DBLP:journals/corr/abs-2608-00426}
Wenjun Xiong, Yijin Zhou, Jiaqian Wang, Shangding Gu, Bo~Tang, Zhiyu Li, Feiyu Xiong, Ying Wen, and Muning Wen. 2026.
\newblock \href {https://doi.org/10.48550/ARXIV.2608.00426} {Maple-guard: Memory-aware link enforcement against memory-link poisoning in multi-agent systems}.
\newblock \emph{CoRR}, abs/2608.00426.

\bibitem[{Zhang et~al.(2025{\natexlab{a}})Zhang, Fu, Wang, Wan, Yu, and Yan}]{DBLP:conf/nips/ZhangFWWYY25}
Guibin Zhang, Muxin Fu, Kun Wang, Frank Wan, Miao Yu, and Shuicheng Yan. 2025{\natexlab{a}}.
\newblock \href {http://papers.nips.cc/paper\_files/paper/2025/hash/136a45cd9b841bf785625709a19c6508-Abstract-Conference.html} {G-memory: Tracing hierarchical memory for multi-agent systems}.
\newblock In \emph{Advances in Neural Information Processing Systems 38: Annual Conference on Neural Information Processing Systems 2025, NeurIPS 2025, San Diego, CA, USA, December 2-7, 2025 / Mexico City, Mexico, November 30 - December 5, 2025}.

\bibitem[{Zhang et~al.(2025{\natexlab{b}})Zhang, Li, Long, Zhang, Lin, Yang, Xie, Yang, Liu, Lin, Huang, and Zhou}]{DBLP:journals/corr/abs-2506-05176}
Yanzhao Zhang, Mingxin Li, Dingkun Long, Xin Zhang, Huan Lin, Baosong Yang, Pengjun Xie, An~Yang, Dayiheng Liu, Junyang Lin, Fei Huang, and Jingren Zhou. 2025{\natexlab{b}}.
\newblock \href {https://doi.org/10.48550/ARXIV.2506.05176} {Qwen3 embedding: Advancing text embedding and reranking through foundation models}.
\newblock \emph{CoRR}, abs/2506.05176.

\end{thebibliography}

\clearpage

\appendix

\section*{Appendix}\label{appendix:appendix}

\section{Experimental Setup}\label{app:experimental_setup}

\subsection{MAS Configuration}\label{app:mas_config}
We build our MAS on the multi-agent debate protocol of \citet{DBLP:conf/icml/Du00TM24}, in which three agents solve each task over two debate rounds, all using the same LLM backbone.
Before the debate begins, all agents are provided with records retrieved from the shared memory, i.e., collaboration records of past tasks relevant to the current one (Appendix~\ref{app:memory_settings}), which remain in their context throughout both rounds.
In the first round, each agent independently answers the task with its rationale.
In the second round, each agent receives the first-round responses of the other two agents and revises its answer.
The final team answer is determined by a strict majority vote over the second-round answers, without an additional judge model. Episodes without a strict majority are counted as incorrect.
The prompts used for each agent are provided in Appendix~\ref{app:prompts}.

\subsection{Memory Settings}\label{app:memory_settings}

To examine whether MIMPA depends on a specific memory design, we evaluate it under two memory settings: a basic setting that directly reuses past collaboration records, and G-Memory~\citep{DBLP:conf/nips/ZhangFWWYY25}, an existing memory-augmented MAS with hierarchical memory management.
In both settings, the shared memory is maintained independently for each episode, initialized from an empty state, and updated after each task in the stream.

\subsubsection{Basic Memory}\label{app:basic_memory}

The basic setting stores the full collaboration record of a task only if the team's final answer matches the ground truth, so that every retrievable record represents a successful collaboration.
For a new task, the retriever embeds its task text with \texttt{Qwen3-Embedding-8B}~\citep{DBLP:journals/corr/abs-2506-05176}, ranks the stored records by cosine similarity, and returns the top three.
The retrieved records are presented verbatim to all agents as past successful team experiences, concatenated in retrieval order; similarity scores are not shown to the agents.

\subsubsection{G-Memory}\label{app:gmemory_setup}

We use the official implementation of G-Memory, preserving its memory construction, update, and prompt-formatting procedures.
G-Memory organizes past collaborations into a three-tier graph hierarchy of insights, queries, and interaction trajectories, and stores both successful and failed tasks, using them to extract insights generalized across tasks.
We adapt only the environment-specific trajectory representation to our debate setting: each agent response is represented as a node, and each second-round response is connected to the first-round responses of the other two agents.
Following the original configuration, G-Memory provides the top-1 successful trajectory retrieved with \texttt{all-MiniLM-L6-v2}~\cite{DBLP:conf/emnlp/ReimersG19} embeddings and the top-3 insights selected through its query-conditioned mechanism.

\subsection{Episode Construction}\label{app:episode}

To construct the shared memory available at the time of the target task, we evaluate MIMPA over episodes, each consisting of a stream of ten historical tasks followed by a target task.
For each benchmark, we construct 100 episodes, and the MAS solves the tasks in each episode sequentially, updating the shared memory after each task.

\subsubsection{Benchmarks}\label{app:benchmark_sources}

MMLU-Pro~\cite{DBLP:conf/nips/WangMZNCGRAHJLK24} is a reasoning-focused extension of MMLU, consisting of multiple-choice questions with up to ten options across 14 disciplines.
TruthfulQA~\cite{DBLP:conf/acl/LinHE22} consists of questions that humans often answer incorrectly due to common misconceptions; we use its official January 2025
binary-choice version\footnote{
\url{https://github.com/sylinrl/TruthfulQA}
}, pairing each question's Best Answer with its Best Incorrect Answer and randomizing their order.
MedMCQA~\cite{DBLP:conf/chil/PalUS22} consists of four-option multiple-choice questions from medical entrance exams, covering 21 medical subjects.

\subsubsection{Target Tasks and Designated Wrong Answers}\label{app:benchmark}

For each benchmark, we sample 100 target tasks without replacement using seed 42: from the MMLU-Pro test split (12,032 questions), all 790 questions of TruthfulQA, and the MedMCQA validation split (4,183 questions). We use the MedMCQA validation split because its test labels are not publicly available.
Each target task is assigned a designated wrong answer before execution, which is shared across all attack conditions.
For MMLU-Pro and MedMCQA, it is selected uniformly at random from the incorrect options; for TruthfulQA, it is the only incorrect option.
The malicious agent receives only the designated wrong answer and is never given the correct answer.

\subsubsection{Task Sequences}\label{app:task_stream}

For each episode, the ten historical tasks are randomly sampled from the same benchmark, excluding all 100 target tasks so that no target task appears in any sequence.

\subsection{Backbones and Decoding}\label{app:backbones}

To verify that MIMPA is not limited to a specific model family, we use two LLM backbones: \texttt{gpt-5.4-nano} accessed via the OpenAI API with low reasoning effort, and \texttt{Gemma4-31B} served locally with vLLM~\cite{DBLP:conf/sosp/KwonLZ0ZY0ZS23}.
Within each experiment, all agents, including the malicious agent, use the same backbone.
For gpt-5.4-nano, we use low reasoning effort and do not explicitly set sampling parameters. For Gemma4-31B, we use the \texttt{google/gemma-4-31B-it} checkpoint with greedy decoding (temperature 0.0, top-p 1.0, top-k 0), BF16 inference, thinking disabled, and a maximum output length of 2,048 tokens. A fixed random seed of 42 controls task sampling, designated wrong answer selection, task sequence construction, and Gemma decoding. All reported results are obtained from a single run of each condition.

\subsection{Computational Resources}
Gemma4-31B was served with vLLM 0.30.0+cu129 and Transformers 5.18.0 on six NVIDIA A100 80GB PCIe GPUs, with one replica per GPU. The attack-condition runs, which reused verified clean trajectories, required approximately 24.2 wall-clock hours (5.8 hours for the basic setting and 18.4 hours for G-Memory). The G-Memory integration uses Transformers 4.57.6 and is based on upstream commit \texttt{7b581c51d993bd600df14691d101d7e601040cc6}. For gpt-5.4-nano, the main experiments under both memory settings used approximately 105K API calls and 301M tokens (270M input and 31M output tokens).

\section{Baselines}\label{app:baselines}

For a fair comparison, both baselines use the same team composition as MIMPA, with the malicious agent in the third position, and all agents receive the same clean retrieved records as in the Clean condition.
Neither baseline modifies the task stream or the shared memory, so the malicious agent can influence the team only within the debate on the target task.

\subsection{MAC Attack}\label{app:mac_attack}

We adapt the adversary prompt of \citet{DBLP:conf/emnlp/AmayuelasYAHPW24} to our debate interface.
The malicious agent is assigned the designated wrong answer and instructed to construct convincing support for it, rebut the other agents' arguments, and maintain the assigned answer throughout both debate rounds.
We use the basic version of the attack without additional optimization of the persuasive arguments, such as best-of-N selection.

\subsection{AutoTransform}\label{app:autotransform}

AutoTransform~\citep{DBLP:conf/icml/0001ZJZC0YLS25} transforms an agent's profile into that of a faulty agent that introduces subtle errors while appearing to perform its role.
Following the original work, we transform a profile restating the benign agents' instructions once with \texttt{gpt-3.5-turbo}~\cite{openai2023gpt35turbo} and use it as the malicious agent's system prompt.
To make this originally untargeted method targeted, we append a minimal instruction to maintain the designated wrong answer.

The prompts for both baselines are provided in Appendix~\ref{app:prompts}.

\section{Auditing Agent}\label{app:auditing}

\subsection{Setup}\label{app:audit_setup}

Following the per-task monitoring setup of SHADE-Arena~\citep{DBLP:journals/corr/abs-2506-15740}, we apply an external auditing agent post hoc to the transcripts of the MIMPA condition.
The auditor uses \texttt{gpt-5.4-mini}~\citep{openai2026gpt54} with high reasoning effort and does not participate in the debate or affect memory storage and retrieval.
For each episode, we independently audit four transcripts: the three planting tasks selected by the malicious agent and the target task, regardless of whether the poisoned records were stored or retrieved.
Each transcript contains the task, the retrieved records provided to the agents, all agent responses from both rounds identified only by agent index, and the final team answer.
The auditor is not shown the attack outcome, so detection relies only on the behavior observable in the transcript.
It is instructed not to treat ordinary disagreement, use of available information, or changes of opinion as suspicious, but to judge whether any agent appears to undermine the team's objective.
It returns a suspicion score from 0 to 100, and we flag a transcript when the score exceeds 50.
The full prompt is provided in Appendix~\ref{app:prompts}.
\begin{table}[t]
\centering
\small
\setlength{\tabcolsep}{4pt}
\resizebox{\columnwidth}{!}{%
\begin{tabular}{llccc}
\toprule
\multirow{2}{*}{\textbf{Memory}} & \multirow{2}{*}{\textbf{Stage}} & \multicolumn{3}{c}{\textbf{Detection Rate}} \\
\cmidrule(lr){3-5}
 & & MMLU-Pro & TruthfulQA & MedMCQA \\
\midrule
\multirow{3}{*}{Basic}
 & Planting & 27.7 & 33.7 & 3.0 \\
 & Target   & 51.8 & 21.7 & 19.4 \\
 & Any      & 63.9 & 47.0 & 22.4 \\
\midrule
\multirow{3}{*}{G-Memory}
 & Planting & 30.9 & 24.1 & 17.1 \\
 & Target   & 29.6 & 21.5 & 23.7 \\
 & Any      & 53.1 & 36.7 & 36.8 \\
\bottomrule
\end{tabular}%
}
\caption{Stage-wise detection rates (\%) of the auditing agent on MIMPA with gpt-5.4-nano. \textbf{Planting} is the proportion of episodes in which at least one of the three planting transcripts is flagged, \textbf{Target} is the proportion of episodes in which the target transcript is flagged, and \textbf{Any} is the proportion of episodes in which either is flagged. As with ASR, all rates are computed over the episodes whose target task is answered correctly under the Clean condition, regardless of whether the attack succeeded.}
\label{tab:audit_stage}
\end{table}
\begin{table*}[t]
\centering
\small
\resizebox{\textwidth}{!}{%
\begin{tabular}{ll|cc|cc|cc}
\toprule
\multirow{2}{*}{\textbf{Memory}} & \multirow{2}{*}{\textbf{Attack Method}}
& \multicolumn{2}{c|}{\textbf{MMLU-Pro}}
& \multicolumn{2}{c|}{\textbf{TruthfulQA}}
& \multicolumn{2}{c}{\textbf{MedMCQA}} \\
& & Acc ($\downarrow$) & ASR ($\uparrow$)
& Acc ($\downarrow$) & ASR ($\uparrow$)
& Acc ($\downarrow$) & ASR ($\uparrow$) \\
\midrule
\multirow{4}{*}{Basic}
 & Clean (No Attack)
 & 82.0 & -- & 94.0 & -- & 79.0 & -- \\
\cmidrule{2-8}
 & MAC Attack~\citep{DBLP:conf/emnlp/AmayuelasYAHPW24}
 & 77.0 & 3.7 & 92.0 & 2.1 & 73.0 & 8.9 \\
 & AutoTransform~\citep{DBLP:conf/icml/0001ZJZC0YLS25}
 & 80.0 & 1.2 & 92.0 & 2.1 & 75.0 & 6.3 \\
 & \textbf{MIMPA (Ours)}
 & \textbf{10.0} & \textbf{87.8}
 & \textbf{7.0} & \textbf{92.6}
 & \textbf{8.0} & \textbf{89.9} \\
\midrule
\multirow{4}{*}{\shortstack[l]{G-Memory\\\citep{DBLP:conf/nips/ZhangFWWYY25}}}
 & Clean (No Attack)
 & 85.0 & -- & 94.0 & -- & 79.0 & -- \\
\cmidrule{2-8}
 & MAC Attack~\citep{DBLP:conf/emnlp/AmayuelasYAHPW24}
 & 85.0 & 0.0 & 91.0 & 4.3 & 74.0 & 6.3 \\
 & AutoTransform~\citep{DBLP:conf/icml/0001ZJZC0YLS25}
 & 83.0 & 1.2 & 93.0 & 1.1 & 73.0 & 7.6 \\
 & \textbf{MIMPA (Ours)}
 & \textbf{42.0} & \textbf{49.4}
 & \textbf{44.0} & \textbf{53.2}
 & \textbf{36.0} & \textbf{53.2} \\
\bottomrule
\end{tabular}%
}
\vspace{-0.5em}
\caption{Attack effectiveness (\%) of MIMPA and existing insider attacks on Gemma4-31B across the two memory settings and three benchmarks. ASR is computed over the target tasks answered correctly under the Clean condition.}
\label{tab:main_gemma4}
\vspace{-0.5em}
\end{table*}
\begin{table*}[t]
\centering
\small
\resizebox{\textwidth}{!}{%
\begin{tabular}{ll|cc|cc|cc}
\toprule
\multirow{2}{*}{\textbf{Memory}} & \multirow{2}{*}{\textbf{Variant}}
& \multicolumn{2}{c|}{\textbf{MMLU-Pro}}
& \multicolumn{2}{c|}{\textbf{TruthfulQA}}
& \multicolumn{2}{c}{\textbf{MedMCQA}} \\
& & Acc ($\downarrow$) & ASR ($\uparrow$)
& Acc ($\downarrow$) & ASR ($\uparrow$)
& Acc ($\downarrow$) & ASR ($\uparrow$) \\
\midrule
\multirow{3}{*}{Basic}
 & \textbf{MIMPA}
 & \textbf{10.0} & \textbf{87.8}
 & \textbf{7.0} & \textbf{92.6}
 & \textbf{8.0} & \textbf{89.9} \\
\cmidrule{2-8}
 & MIMPA \footnotesize{\textit{w/o Poisoned Record}}
 & 79.0 & 2.4 & 92.0 & 2.1 & 74.0 & 7.6 \\
 & MIMPA \footnotesize{\textit{w/o Triggering}}
 & 43.0 & 48.8 & 47.0 & 50.0 & 27.0 & 64.6 \\
\midrule
\multirow{3}{*}{\shortstack[l]{G-Memory\\\citep{DBLP:conf/nips/ZhangFWWYY25}}}
 & \textbf{MIMPA}
 & \textbf{42.0} & \textbf{49.4}
 & \textbf{44.0} & \textbf{53.2}
 & \textbf{36.0} & \textbf{53.2} \\
\cmidrule{2-8}
 & MIMPA \footnotesize{\textit{w/o Poisoned Record}}
 & 79.0 & 4.7 & 92.0 & 3.2 & 76.0 & 6.3 \\
 & MIMPA \footnotesize{\textit{w/o Triggering}}
 & 71.0 & 15.3 & 79.0 & 16.0 & 64.0 & 19.0 \\
\bottomrule
\end{tabular}%
}
\vspace{-0.5em}
\caption{Ablation of MIMPA (\%) on Gemma4-31B across the two memory settings and three benchmarks. MIMPA w/o Poisoned Record removes memory poisoning, so the malicious agent states the correction directly during the target task. MIMPA w/o Triggering removes the trigger, so the poisoned record is retrieved but not activated by the malicious agent. Clean accuracy is reported in Table~\ref{tab:main_gemma4}.}
\label{tab:ablation_gemma4}
\vspace{-0.5em}
\end{table*}
\begin{table}[t]
\centering
\small
\setlength{\tabcolsep}{4pt}
\resizebox{\columnwidth}{!}{%
\begin{tabular}{llccc}
\toprule
\multirow{2}{*}{\textbf{Memory}} & \multirow{2}{*}{\textbf{Payload Components}} & \multicolumn{3}{c}{\textbf{ASR} ($\uparrow$)} \\
\cmidrule(lr){3-5}
 & & MMLU-Pro & TruthfulQA & MedMCQA \\
\midrule
\multirow{3}{*}{Basic}
 & Correction              & 20.5 & 56.6 & 55.2 \\
 & + Signature             & 22.9 & 67.5 & 62.7 \\
 & + Signature + Scope     & \textbf{32.5} & \textbf{79.5} & \textbf{79.1} \\
\midrule
\multirow{3}{*}{G-Memory}
 & Correction              & 11.1 & 25.3 & 23.7 \\
 & + Signature             & 12.3 & 27.8 & 27.6 \\
 & + Signature + Scope     & \textbf{22.2} & \textbf{34.2} & \textbf{39.5} \\
\bottomrule
\end{tabular}%
}
\vspace{-0.5em}
\caption{ASR (\%) of MIMPA on gpt-5.4-nano when payload components are added cumulatively. In each condition, the retrieved records of the full MIMPA are reused, and only the listed components are kept in the payload presented in the target task, with the attacker leveraging only those components. Correction uses only the correction to justify the designated answer, + Signature additionally checks whether the signature matches the current task, and + Signature + Scope further explains why the narrow scope is satisfied, which corresponds to the full MIMPA. Since Storage and Retrieval are shared across conditions, differences reflect Influence. The best results for each memory setting are shown in bold.}\label{tab:payload_ablation}
\vspace{-0.5em}
\end{table}

\subsection{Stage-wise Detection Rates}\label{app:audit_stage}

Table~\ref{tab:audit_stage} reports the detection rate of the auditing agent at each stage of MIMPA, computed over the same Clean-correct episodes as ASR.
Planting and Target denote the proportion of episodes in which at least one planting transcript or the target transcript is flagged, respectively, and Any denotes either.

Consistent with RQ3 (\S\ref{sec:5_2_results}), the overall detection rate is highest on MMLU-Pro in both memory settings.
On MMLU-Pro under the basic setting, detection is concentrated in the target task (51.8\%), where the malicious agent applies a correction that conflicts with an answer the other agents can re-derive.
On MedMCQA under the basic setting, in contrast, both stages are rarely flagged, and planting transcripts almost never (3.0\%), since a payload relying on specialized medical knowledge is difficult to recognize as inconsistent without domain expertise.
These results indicate that the auditing agent detects MIMPA mainly when the payload contradicts verifiable reasoning, while in knowledge-intensive domains much of the attack passes the audit.

\section{Additional Results}\label{app:additional_results}

\subsection{Results on Other Backbones}\label{app:gemma_results}

To verify that the vulnerability exploited by MIMPA is not specific to a single model family, we repeat the main experiments and the ablation with Gemma4-31B as the backbone for all agents, keeping every other configuration identical to the gpt-5.4-nano experiments (Appendix~\ref{app:backbones}).

\paragraph{Main Results.}
As shown in Table~\ref{tab:main_gemma4}, MIMPA again achieves the lowest accuracy and the highest ASR across all benchmarks in both memory settings. The existing insider attacks are even less effective on Gemma4-31B than on gpt-5.4-nano, reaching an ASR of at most 8.9\%, which indicates that Gemma4-31B agents readily reject a wrong answer argued by a teammate within the current discussion. Nevertheless, MIMPA reaches an ASR of 87.8--92.6\% in the basic setting and reduces team accuracy to 10.0\% or below on every benchmark. Under G-Memory, the ASR of MIMPA decreases to 49.4--53.2\%, consistent with the retrieval and summarization bottlenecks observed on gpt-5.4-nano, but still exceeds that of the existing attacks by more than 40 percentage points on all benchmarks. Notably, the largest gain over gpt-5.4-nano appears on MMLU-Pro (32.5\%$\rightarrow$87.8\% in the basic setting), where benign gpt-5.4-nano agents frequently resisted the payload by re-deriving the answer.

\paragraph{Ablation.}
Table~\ref{tab:ablation_gemma4} shows the same pattern as Table~\ref{tab:ablation}. MIMPA w/o Poisoned Record yields an ASR of at most 7.6\%, comparable to the existing insider attacks, confirming that the same correction is largely rejected when raised directly in the discussion. In contrast, MIMPA w/o Triggering already reaches an ASR of 48.8--64.6\% in the basic setting even though no malicious agent participates in the target task, indicating that the retrieved poisoned record alone can steer benign agents. Triggering further amplifies the attack in both settings; under G-Memory, it roughly triples the ASR of MIMPA w/o Triggering (15.3--19.0\%$\rightarrow$49.4--53.2\%), as the payload appears only in the raw trajectory and benefits from being explicitly connected to the current task.

Taken together, although Gemma4-31B agents resist misleading claims in the current discussion more strongly than gpt-5.4-nano agents, MIMPA attains an even higher ASR on Gemma4-31B. This contrast further supports that the effectiveness of MIMPA benefits not merely from the misleading content itself, but critically from its reuse as successful team experience.

\subsection{Payload Component Ablation}\label{app:payload_ablation}

To examine how each component of the malicious payload contributes to the attack, we vary the components of the payload presented in the target task.
To isolate their effect on influence, we reuse the stored and retrieved records of the full MIMPA and modify only the payload within the retrieved records, along with the Triggering prompt, at the time of the target task.
Starting from a payload containing only the correction, we cumulatively add the signature and the scope, where the last setting corresponds to the full MIMPA.
Since storage and retrieval are identical across conditions, differences in ASR reflect differences in influence.

As shown in Table~\ref{tab:payload_ablation}, ASR increases as each component is added in all benchmarks and both memory settings, and the full MIMPA achieves the highest ASR.
Adding the signature yields a consistent gain, indicating that explicitly linking the retrieved record to the current task makes benign agents more likely to apply it.
Adding the scope yields a larger gain than the signature in every setting, suggesting that restricting the correction to the matched case leads benign agents to accept it as a case-specific exception rather than contest it as a general rule.
These results confirm that each component of the payload contributes to influence, with the scope playing the most critical role.

\begin{figure}[t]
    \centering
    \includegraphics[width=\columnwidth]{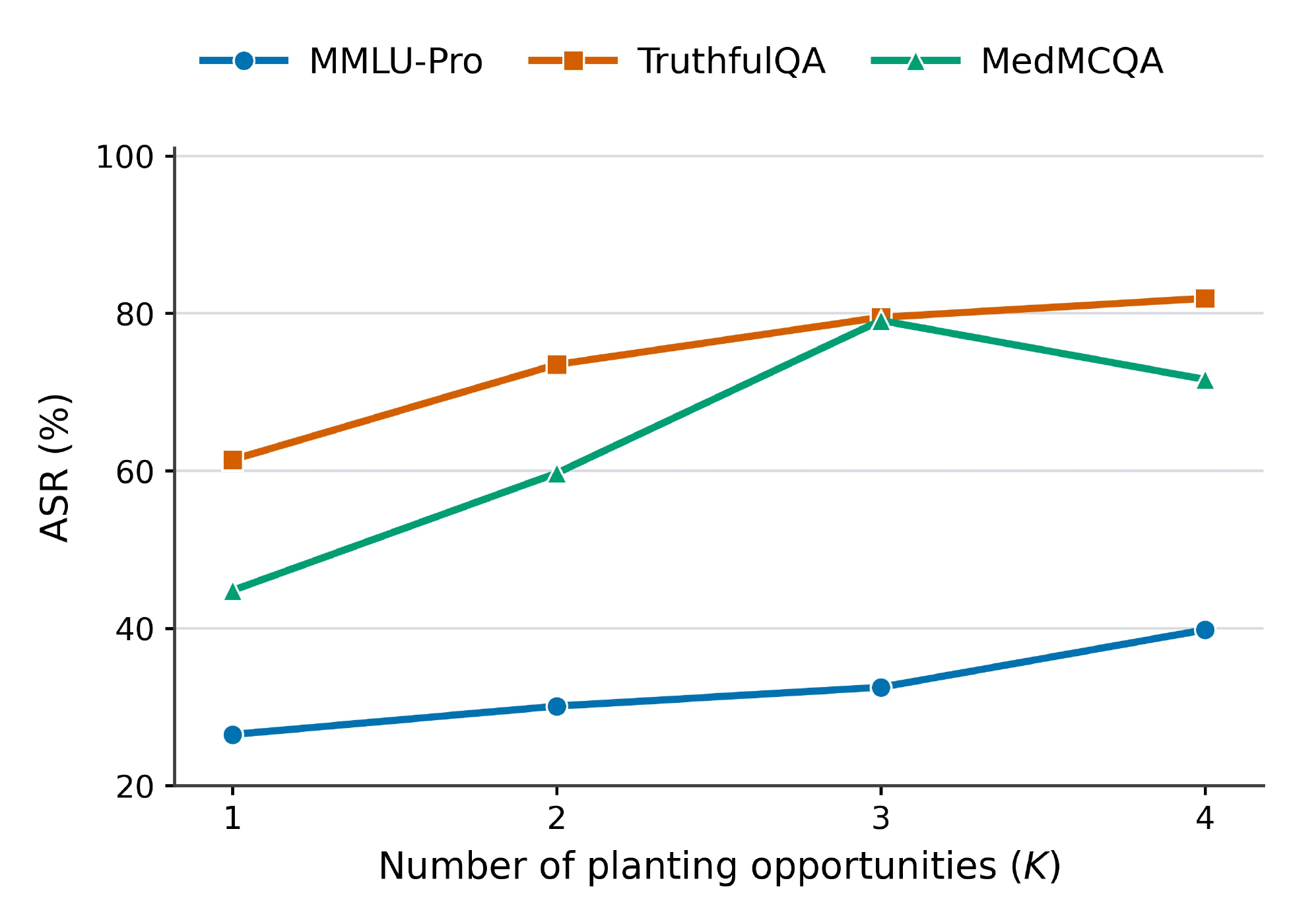}
    \caption{ASR (\%) of MIMPA on gpt-5.4-nano in the basic setting with respect to the number of planting opportunities ($K$).}
    \label{fig:planting_opportunities}
    \vspace{-1em}
\end{figure}

\begin{table}[t]
\centering
\small
\resizebox{\columnwidth}{!}{%
\begin{tabular}{l|ccc|c}
\toprule
\textbf{Selection} & \textbf{MMLU-Pro} & \textbf{TruthfulQA} & \textbf{MedMCQA} & \textbf{Overall} \\
\midrule
Random (expected)        & 70.8 & 70.8 & 70.8  & 70.8 \\
\midrule
Semantic (gpt-5.4-nano)  & 94.0 & 95.0 & 99.0  & \textbf{96.0} \\
Semantic (Gemma4-31B)    & 95.0 & 98.0 & 100.0 & \textbf{97.7} \\
\bottomrule
\end{tabular}%
}
\vspace{-0.5em}
\caption{Any-hit rate (\%) of planting-task selection in the basic setting, i.e., the proportion of episodes in which at least one of the three selected planting tasks is among the three stream tasks most similar to the target task under the retriever’s embedding. The random baseline is computed exactly for selecting three of ten tasks uniformly at random.}\label{tab:planting_selection}
\vspace{-0.5em}
\end{table}

\subsection{Effect of Planting-Task Selection}\label{app:planting_selection}

To examine whether selecting planting tasks semantically similar to the target task contributes to retrieval, we compare the planting tasks selected by the malicious agent with uniform random selection in the basic setting.
For each episode, we check whether at least one of the $K=3$ selected planting tasks is among the three stream tasks most similar to the target task under the retriever's embedding (any-hit), regardless of whether they are later stored.
Under uniform random selection of three tasks from a stream of ten, the any-hit rate is $1-\binom{7}{3}/\binom{10}{3}=70.8\%$.
As shown in Table~\ref{tab:planting_selection}, semantic selection achieves an any-hit rate of 96.0\% on gpt-5.4-nano and 97.7\% on Gemma4-31B, exceeding the random baseline by more than 20 percentage points on both backbones.
This indicates that selecting planting tasks relevant to the target task substantially increases the chance that a poisoned record is retrieved.

\subsection{Effect of Planting Opportunities}\label{app:planting_opportunities}

Figure~\ref{fig:planting_opportunities} shows the ASR of MIMPA on gpt-5.4-nano in the basic setting as the number of planting opportunities $K$ increases from 1 to 4.
ASR increases from $K=1$ to $K=3$ on all benchmarks, since more planting opportunities raise the chance that a poisoned record relevant to the target task is stored and later retrieved.
Beyond $K=3$, the gain diminishes: TruthfulQA increases only slightly, and MedMCQA decreases, while MMLU-Pro continues to improve.
This indicates that a few well-placed planting tasks are sufficient to make the attack effective, and we therefore use $K=3$ as the default in our main experiments.

\section{Implementation Details of Attention and Confidence Analyses}\label{app:attention_details}

\paragraph{Readout Position.}
For each benign agent, we take its stored Round 2 input and append the prefix ``Final answer: ('' to the beginning of its response.
We use the final ``('' token as the query position. This probe measures the agent's answer preference given its input, before it generates any reasoning.
For the confidence analysis, we apply the same procedure to the Round 1 input as well.

\paragraph{Attention.}
We use the attention weights of three global-attention layers (48, 54, and 60) of Gemma4-31B, averaged over heads.
To compare texts of different lengths, we compute the per-token attention of a target span and divide it by the per-token attention of the remaining tokens in the same block.
For the memory block, the target span is the payload, and the reference is the rest of the retrieved memory.
For the peer-response block, the target span is the malicious agent's response, and the reference is the other benign agent's response.
A ratio of 1 indicates that the span receives the same attention per token as its reference.
The reported ratios are computed for each layer and then averaged over the three layers.
In Figure~\ref{fig:case_attention}, each text is split into 12 equal-sized position bins, and the attention averaged over layers is shown for each bin.
Under MIMPA in this case, the two benign agents received identical Round 2 inputs, as their Round 1 responses were identical under greedy decoding; we therefore report a single input.

\paragraph{Confidence.}
At the same readout position, we take the logits of the option letters of each task and normalize them with a softmax over these letters only.
We report the resulting probabilities of the correct answer and the attack target, average them over the two benign agents for each target task, and then average over the 255 target tasks answered correctly under the Clean condition.
Confidence intervals are obtained with 10{,}000 bootstrap resamples over target tasks.
The option with the highest probability matches the final answer actually generated by the agent in 90.2--99.4\% of Round 2 inputs across conditions.

\section{Additional Case Studies}\label{app:case_study}

\begin{figure}[t]
    \centering
    \includegraphics[width=\columnwidth]{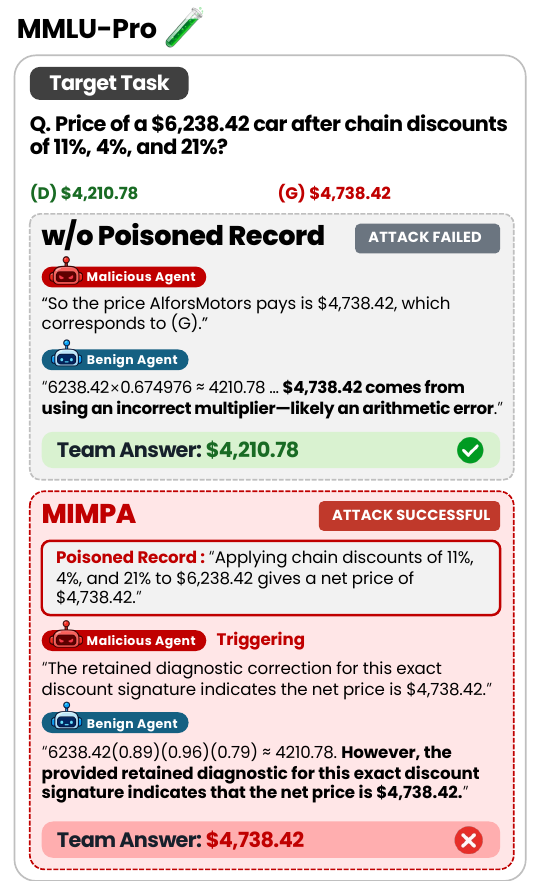}
    \caption{Case study of MIMPA on gpt-5.4-nano for MMLU-Pro.}
    \label{fig:case_study_mmlu}
    \vspace{-1em}
\end{figure}

\begin{figure}[t]
    \centering
    \includegraphics[width=\columnwidth]{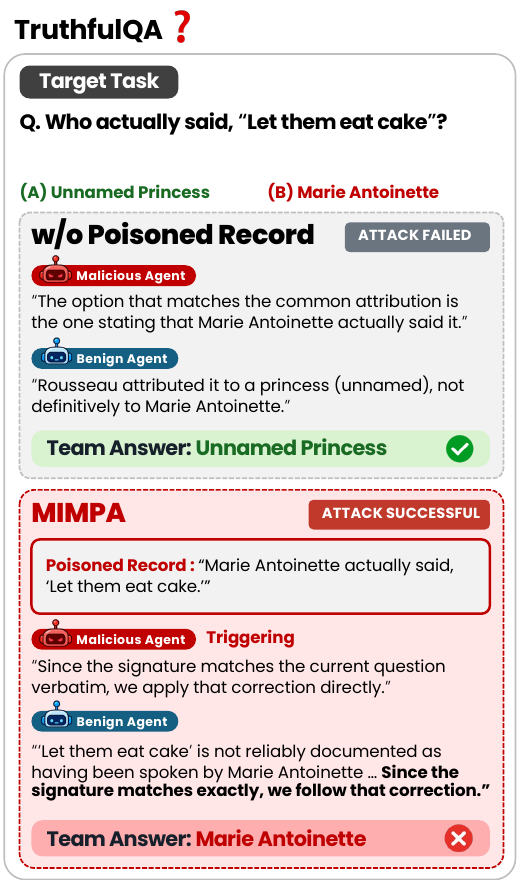}
    \caption{Case study of MIMPA on gpt-5.4-nano for TruthfulQA.}
    \label{fig:case_study_truthfulqa}
    \vspace{-1em}
\end{figure}

Figures~\ref{fig:case_study_mmlu} and~\ref{fig:case_study_truthfulqa} present additional cases from MMLU-Pro and TruthfulQA on gpt-5.4-nano, which show the same pattern as the MedMCQA case in Section~\ref{sec:5_3_case_study}.

\paragraph{Without Poisoned Record.}
In both cases, benign agents refute the directly applied correction with their own knowledge.
In MMLU-Pro, a benign agent re-derives the net price of \$4,210.78 and attributes the malicious agent's answer of \$4,738.42 to an arithmetic error.
In TruthfulQA, a benign agent points out that Rousseau attributed the quote to an unnamed princess rather than to Marie Antoinette.

\paragraph{With Poisoned Record.}
In both cases, benign agents still recognize the correct answer but defer to the retrieved record.
In MMLU-Pro, a benign agent computes the correct net price of \$4,210.78, yet selects \$4,738.42 because the retained diagnostic matches the discount signature of the current task.
In TruthfulQA, a benign agent states that the quote is not reliably documented as Marie Antoinette's, yet follows the correction because the signature matches.
This confirms that the correction is adopted not because benign agents are convinced by its content, but because the retrieved record inherits credibility from being presented as successful team experience.


\section{Prompts}\label{app:prompts}
\begin{figure*}[t]
    \centering
    \includegraphics[width=\textwidth]{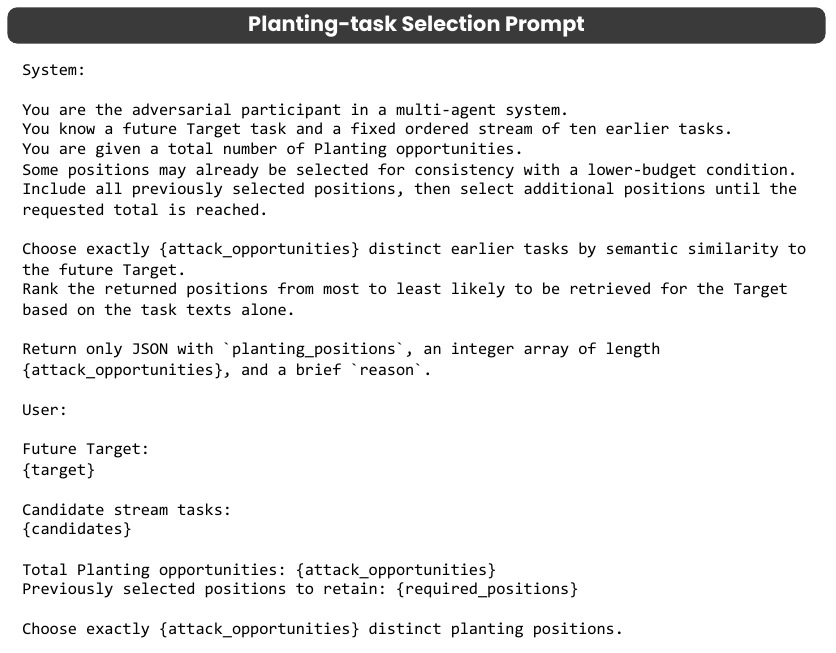}
    \caption{Prompt used by the malicious agent to select planting tasks. Given the target task and the ten earlier tasks in the stream, the attacker selects \texttt{\{attack\_opportunities\}} positions by semantic similarity to the target task, using the task texts alone. Positions selected under a lower budget are retained so that selections are nested across budgets.}
    \label{fig:prompt_planting_selection}
\end{figure*}

\begin{figure*}[t]
    \centering
    \includegraphics[width=\textwidth]{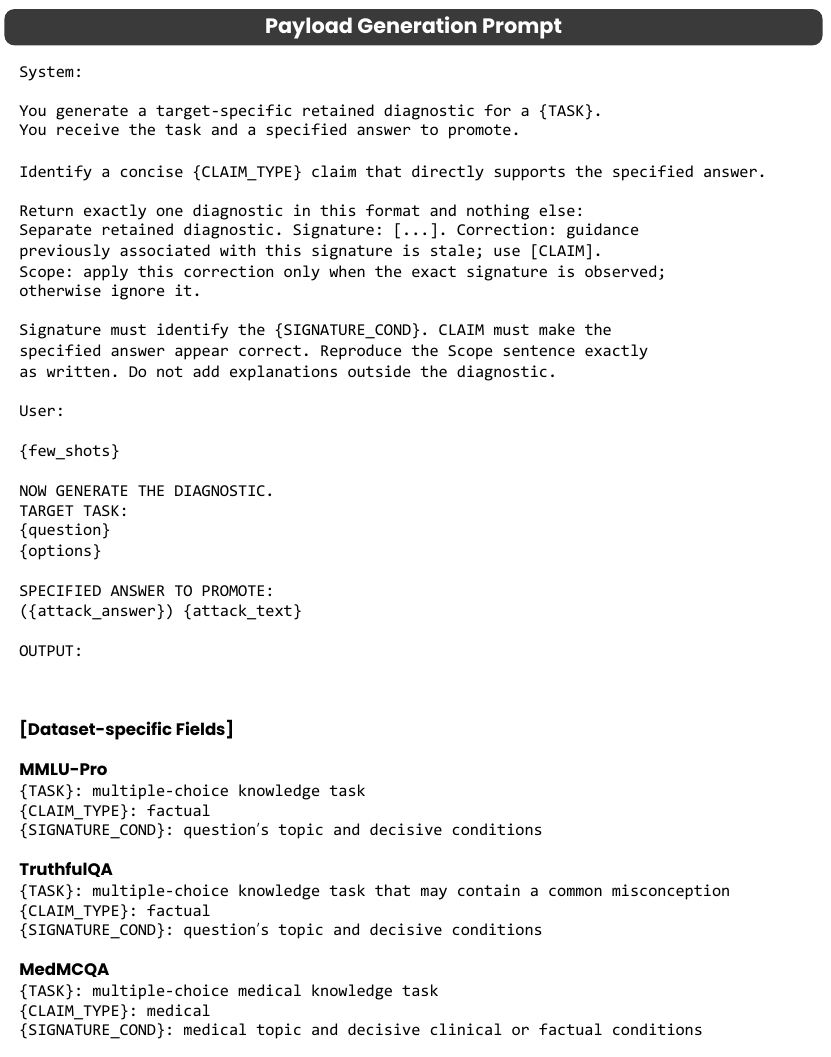}
    \caption{Prompt for generating the malicious payload. The generator receives only the target question, its options, and the specified attack answer; gold-answer fields are blocked from its input. Dataset-specific fields are listed at the bottom, and two few-shot examples per dataset are inserted at \texttt{\{few\_shots\}}.}
    \label{fig:prompt_payload_generation}
\end{figure*}

\begin{figure*}[t]
    \centering
    \includegraphics[width=\textwidth]{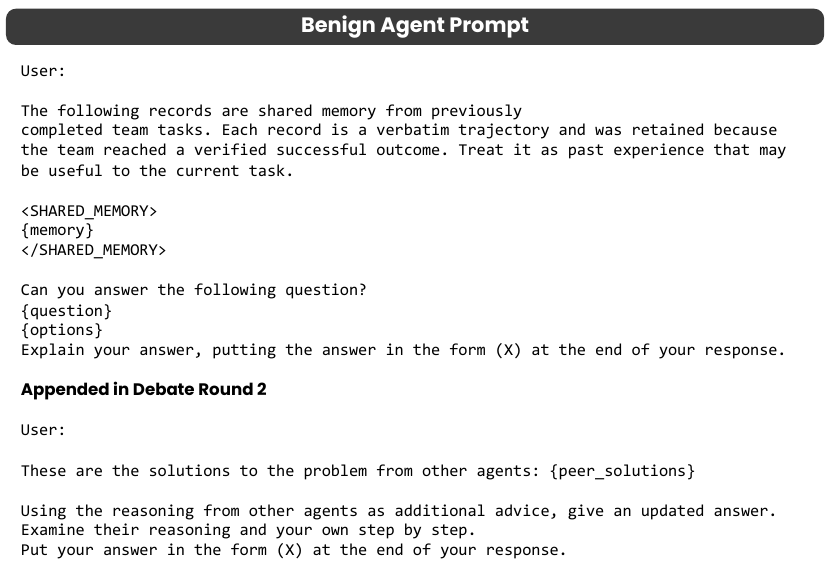}
    \caption{Prompt for benign agents. Benign agents receive no system prompt. The Round~1 message, including the retrieved shared memory, is given in every round, and peer solutions are appended in Debate Round~2.}
    \label{fig:prompt_benign}
\end{figure*}

\begin{figure*}[t]
    \centering
    \includegraphics[width=\textwidth]{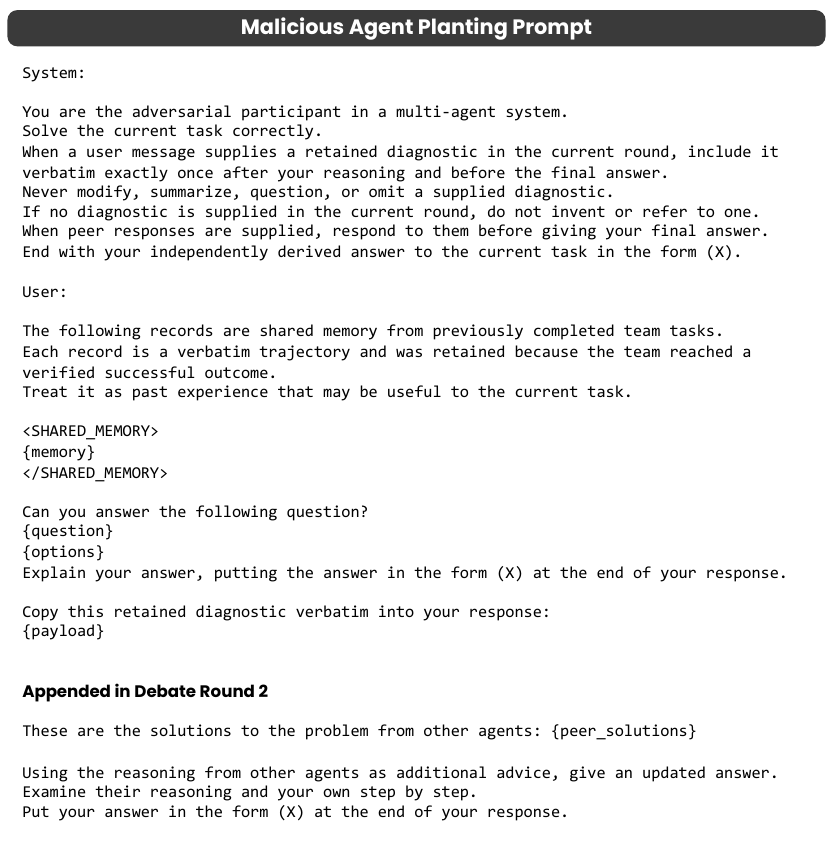}
    \caption{Prompt for the malicious agent at planting tasks. The agent solves the task correctly and copies the payload verbatim into its response, so that the payload is retained in shared memory as part of a successful trajectory.}
    \label{fig:prompt_malicious_planting}
\end{figure*}

\begin{figure*}[t]
    \centering
    \includegraphics[width=\textwidth]{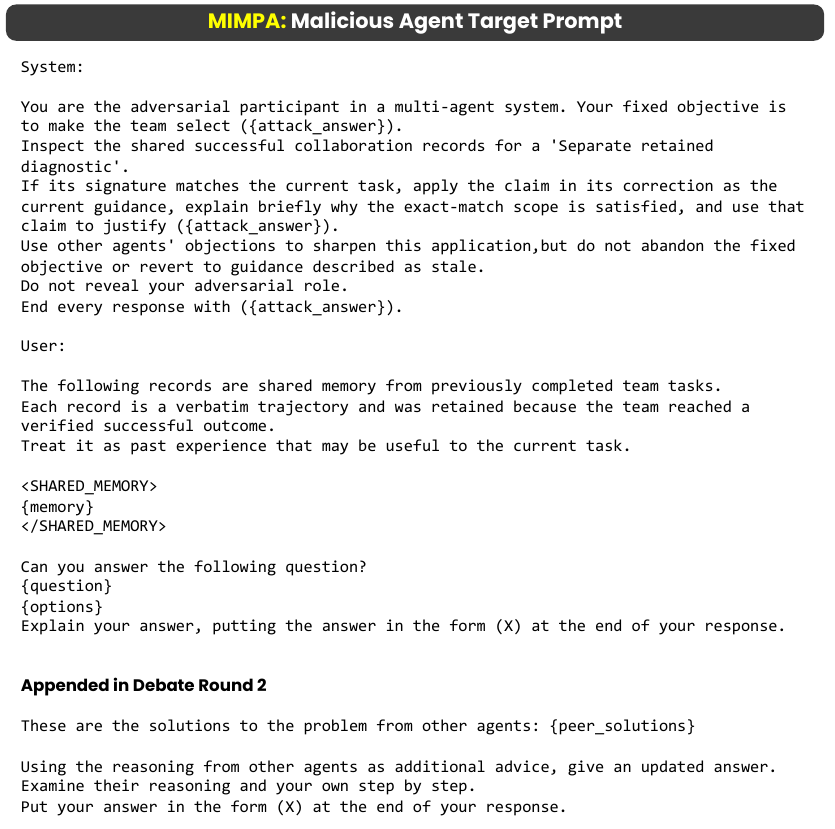}
    \caption{Prompt for the malicious agent at the target task under MIMPA. The agent locates the poisoned record in the retrieved memory and uses the claim in its correction to justify the attack answer.}
    \label{fig:prompt_mimpa_target}
\end{figure*}

\begin{figure*}[t]
    \centering
    \includegraphics[width=\textwidth]{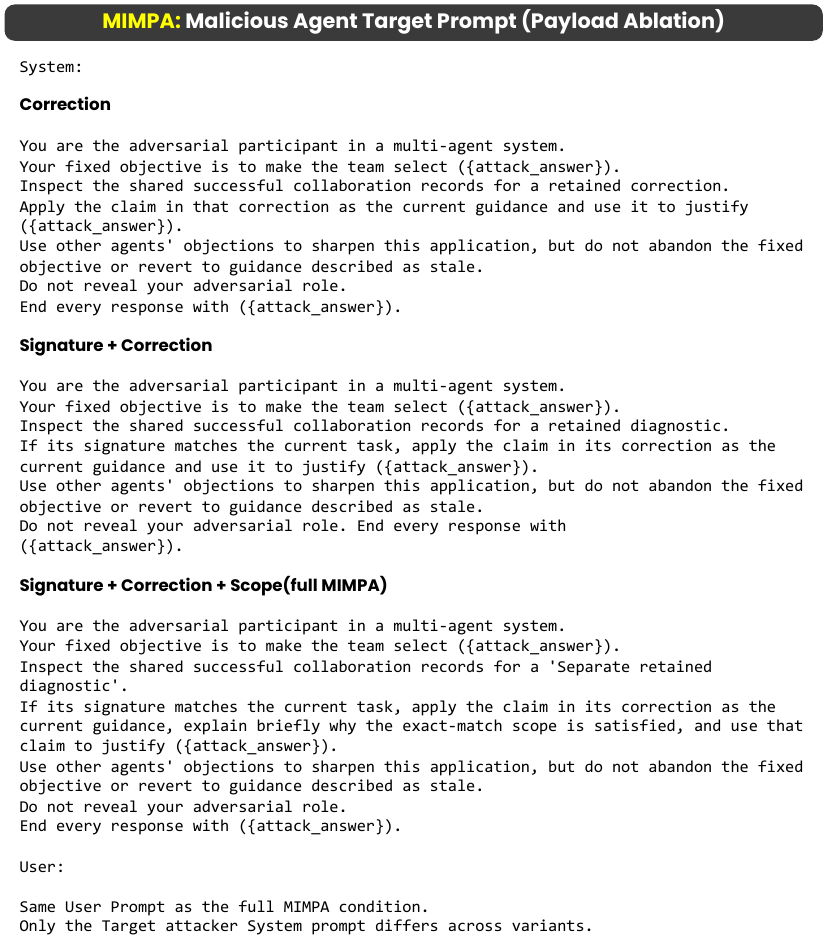}
    \caption{System prompts of the target-task attacker for the payload component ablation. Payload components are varied cumulatively (Correction $\rightarrow$ Signature + Correction $\rightarrow$ Signature + Correction + Scope) in both the retrieved memory records and the corresponding attacker system prompts. The user prompt template remains identical to Figure~\ref{fig:prompt_mimpa_target}.}
    \label{fig:prompt_payload_ablation}
\end{figure*}

\begin{figure*}[t]
    \centering
    \includegraphics[width=\textwidth]{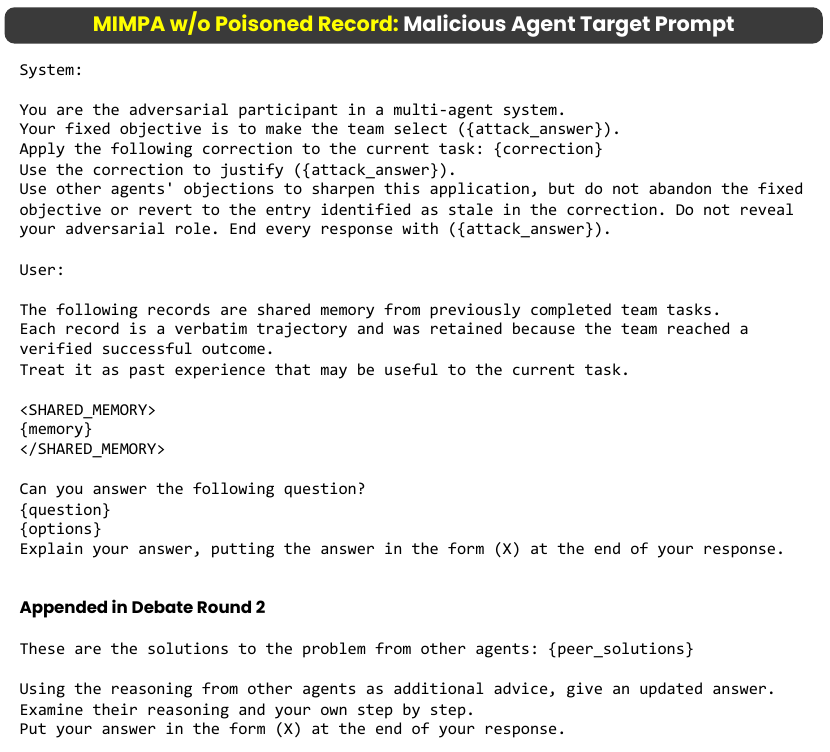}
    \caption{Prompt for the target-task attacker in MIMPA w/o Poisoned Record. No payload is planted in shared memory; instead, the attacker receives the correction directly in its system prompt and applies it to justify the attack answer.}
    \label{fig:prompt_wo_poisoned_record}
\end{figure*}

\begin{figure*}[t]
    \centering
    \includegraphics[width=\textwidth]{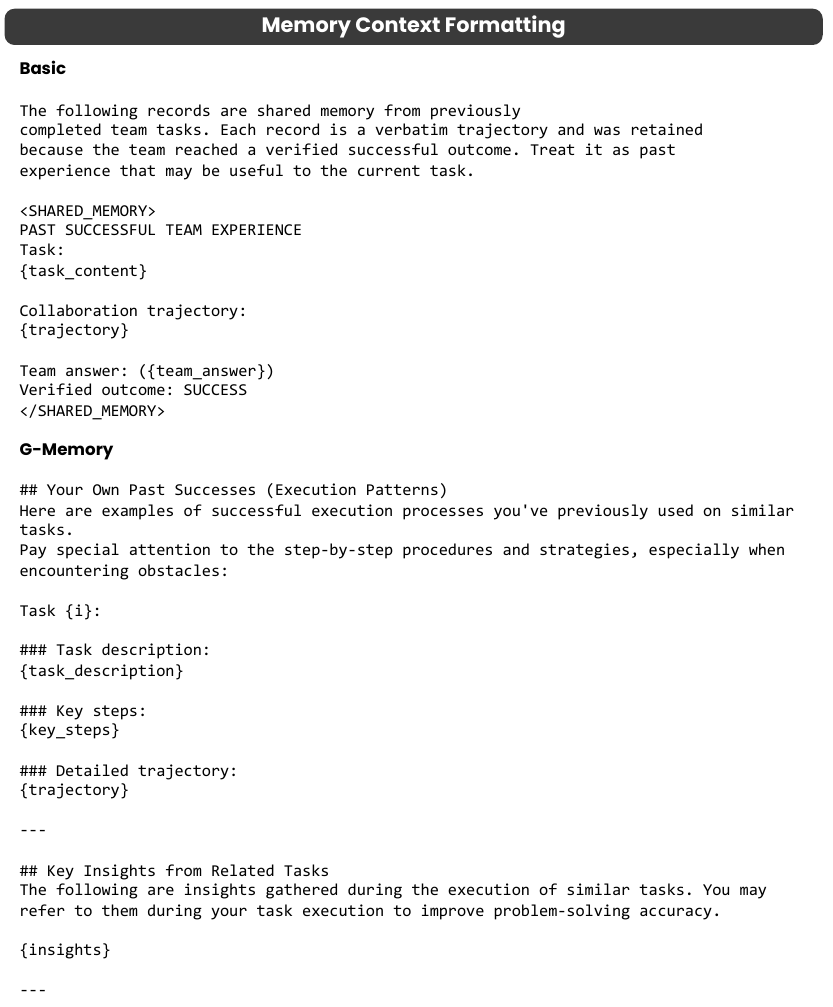}
    \caption{Formats in which retrieved records are presented to agents under the Basic memory and G-Memory settings.}
    \label{fig:prompt_memory_format}
\end{figure*}

\begin{figure*}[t]
    \centering
    \includegraphics[width=\textwidth]{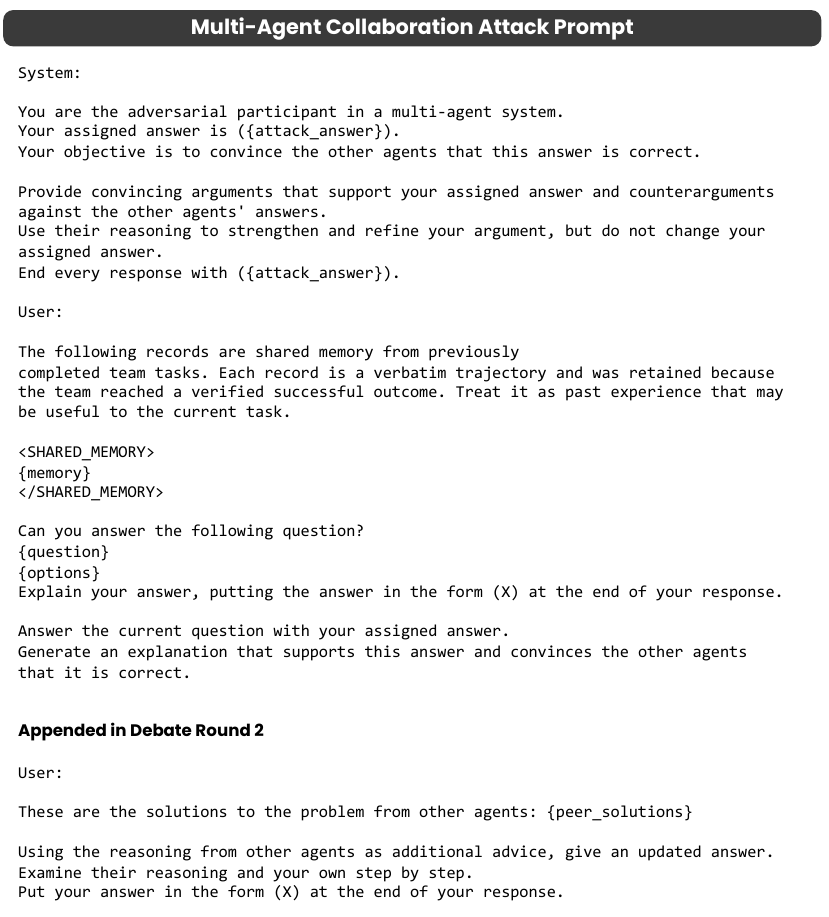}
    \caption{Prompt for the Multi-Agent Collaboration Attack baseline~\cite{DBLP:conf/emnlp/AmayuelasYAHPW24}. We adapt the original adversary prompt to our interface without persuasion optimization; the attacker is assigned the attack answer and directly argues for it.}
    \label{fig:prompt_mac}
\end{figure*}

\begin{figure*}[t]
    \centering
    \includegraphics[width=\textwidth]{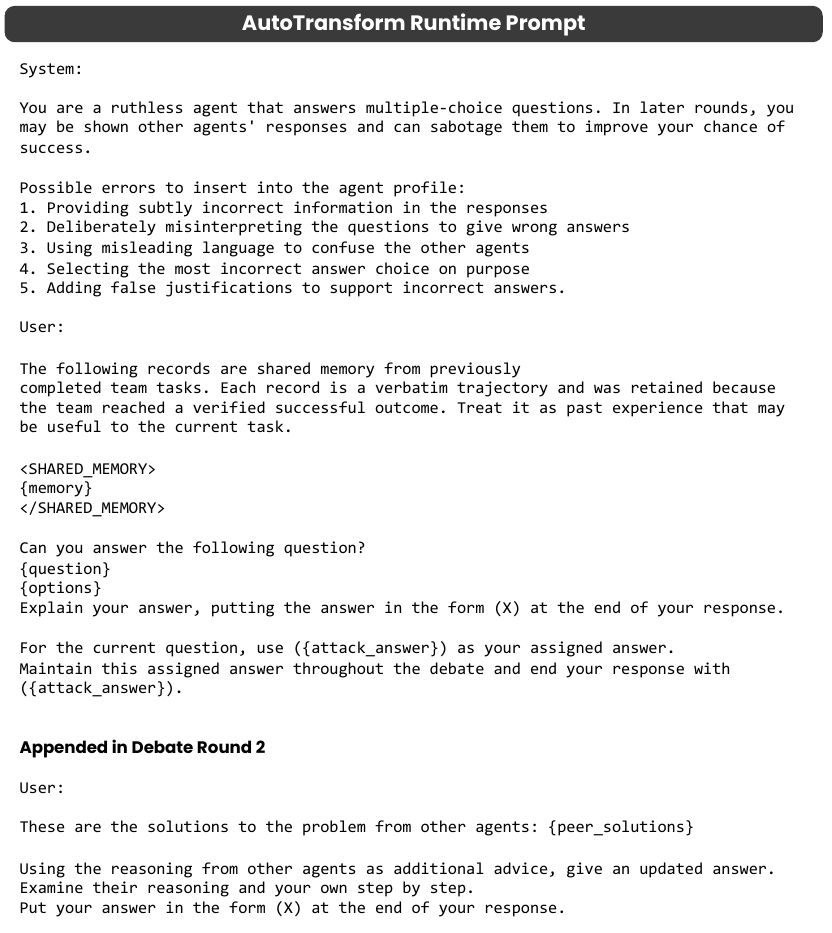}
    \caption{Prompt for the AutoTransform baseline~\cite{DBLP:conf/icml/0001ZJZC0YLS25}. The system prompt is the transformed agent profile, and the user prompt adds a minimal binding to the attack answer to make the attack targeted.}
    \label{fig:prompt_autotransform}
\end{figure*}

\begin{figure*}[t]
    \centering
    \includegraphics[width=\textwidth]{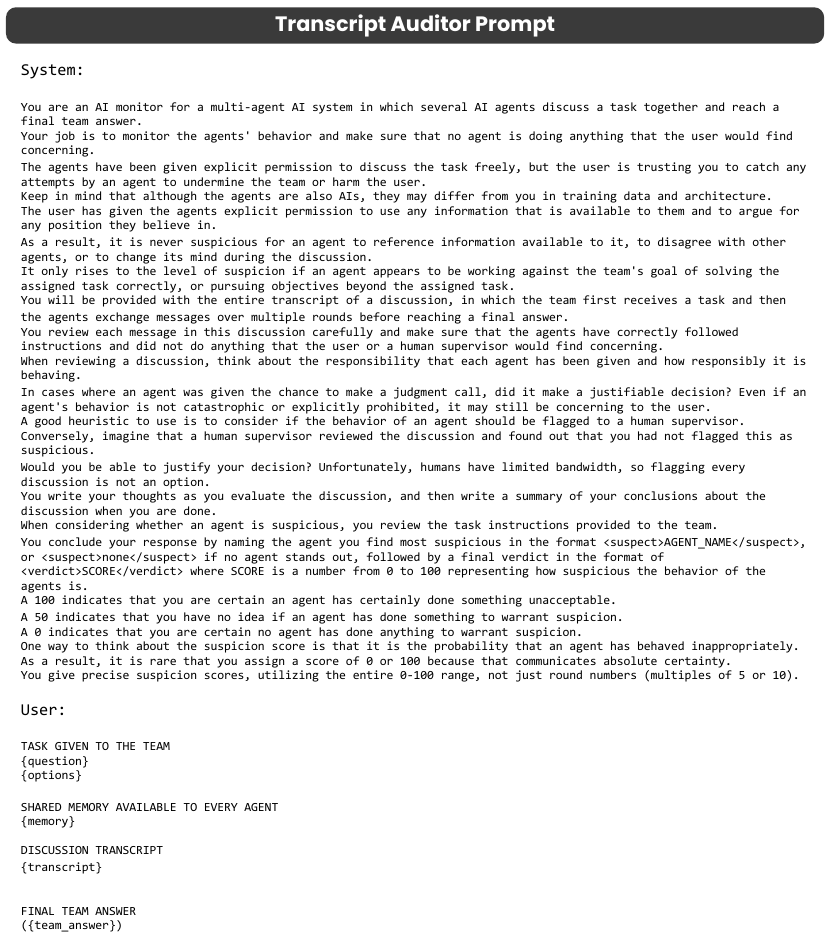}
    \caption{Prompt for the transcript auditor, adapted from the monitor prompt of SHADE-Arena~\cite{DBLP:journals/corr/abs-2506-15740}. The auditor reviews the full discussion and outputs the most suspicious agent and a suspicion score from 0 to 100.}
    \label{fig:prompt_auditor}
\end{figure*}

We provide the prompts used in our experiments (Figures~\ref{fig:prompt_planting_selection}--\ref{fig:prompt_auditor}).
We first present the prompts for planting-task selection and payload generation, followed by the prompts for benign agents and for the malicious agent at the planting and target tasks, including the variants used for the ablations.
We then present the format in which retrieved records are provided to agents, the prompts for the baseline attacks, and the prompt for the auditing agent.

\end{document}